\documentclass[english,aps,manuscript,preprintnumbers,superscriptaddress,eqsecnum,nofootinbib]{revtex4}
\usepackage[T1]{fontenc}
\usepackage[utf8]{luainputenc}
\usepackage{array}
\usepackage{appendix}
\usepackage{multirow}
\usepackage{amsmath}
\allowdisplaybreaks[4]
\usepackage{amssymb}
\usepackage{float}

\usepackage{wasysym}
\usepackage{graphicx}
\usepackage{esint}
\usepackage[marginal]{footmisc}
\numberwithin{equation}{section}
\DeclareMathSizes{12}{12}{7}{7}
\renewcommand{\theequation}{\arabic{section}.\arabic{equation}}

\renewcommand\thesection{\arabic{section}}

\renewcommand{\baselinestretch}{1.1}
\makeatletter

\providecommand{\tabularnewline}{\\}
\newcommand{\lyxdot}{.}

\@ifundefined{textcolor}{}
{%
 \definecolor{BLACK}{gray}{0}
 \definecolor{WHITE}{gray}{1}
 \definecolor{RED}{rgb}{1,0,0}
 \definecolor{GREEN}{rgb}{0,1,0}
 \definecolor{BLUE}{rgb}{0,0,1}
 \definecolor{CYAN}{cmyk}{1,0,0,0}
 \definecolor{MAGENTA}{cmyk}{0,1,0,0}
 \definecolor{YELLOW}{cmyk}{0,0,1,0}
}
\numberwithin{figure}{section}

\makeatother

\usepackage{babel}
\begin{document}
\title{ON THE EVAPORATION OF SOLAR DARK MATTER : SPIN-INDEPENDENT EFFECTIVE OPERATORS}

\author{Zheng-Liang Liang }

\email{liangzl@itp.ac.cn}
\affiliation{Institute of High Energy Physics, Chinese Academy
of Sciences\\19B Yuquan Road, Beijing, 100049, P.R. China}

\author{Yue-Liang Wu}
\email{ylwu@itp.ac.cn}
\affiliation{Kavli Institute for Theoretical Physics China,\\CAS Key Laboratory of Theoretical Physics,\\Institute of Theoretical Physics, Chinese Academy of Sciences\\Zhong Guan Cun Street 55\#, Beijing, 100190, P.R. China}

\author{Zi-Qing Yang }
\email{zqyang@itp.ac.cn}
\affiliation{Kavli Institute for Theoretical Physics China,\\CAS Key Laboratory of Theoretical Physics,\\Institute of Theoretical Physics, Chinese Academy of Sciences\\Zhong Guan Cun Street 55\#, Beijing, 100190, P.R. China}
\author{Yu-Feng Zhou }
\email{yfzhou@itp.ac.cn}
\affiliation{Kavli Institute for Theoretical Physics China,\\CAS Key Laboratory of Theoretical Physics,\\Institute of Theoretical Physics, Chinese Academy of Sciences\\Zhong Guan Cun Street 55\#, Beijing, 100190, P.R. China}

\vspace{3cm}

\begin{abstract}
As a part of the effort to investigate the implications of dark matter
(DM)-nucleon effective interactions on the solar DM detection, in
this paper we focus on the evaporation of the solar DM for a set of
the DM-nucleon spin-independent (SI) effective operators. In order
to put the evaluation of the evaporation rate on a more reliable ground,
we calculate the non-thermal distribution of the solar DM using the
Monte Carlo methods, rather than adopting the Maxwellian approximation.
We then specify relevant signal parameter spaces for the solar DM
detection for various SI effective operators. Based on the analysis,
we determine the minimum DM masses for which the DM-nucleon coupling
strengths can be probed from the  solar neutrino observations.
As an interesting application, our investigation also shows that evaporation effect
can not be neglectd in a recent proposal aiming to solve the solar abundance
problem by invoking the momentum-dependent asymmetric DM in the Sun.

\end{abstract}
\maketitle

\section{\label{sec:Introduction}Introduction}

As the nearest celestial body that is well understood and is capable
of stimulating and responding to the phenomena associated with the
Dark Matter (DM), the Sun is presumed to be an ideal host for the
DM detection. For one thing its deep gravitational well attracts and
traps the Galactic DM particles through the scatter off solar elements,
if there exists a DM-nucleon interaction at the weak scale. For another
thing these captured DM particles may accumulate in the solar core
and subsequently annihilate to primary and secondary high energy neutrino
flux that escape from the dense solar plasma, leaving a smoking-gun
for their presence in the Sun. At present, a number of terrestrial
neutrino detection projects such as IceCube~\cite{Aartsen:2012kia,Albuquerque:2013xna},
Super-Kamiokande~\cite{2011ApJ...742...78T}, Baikal Neutrino Project~\cite{Avrorin:2014swy}
and ANTARES~\cite{Adrian-Martinez:2013ayv} are dedicated to such
observational mission.

In general, the neutrino flux at the detector location is related
to the solar DM annihilation through the following schematic relation:
\begin{equation}
\frac{d\Phi_{\nu}}{dE_{\nu}}=\frac{\Gamma_{A}}{4\pi d_{\odot}^{2}}\frac{dN_{\nu}}{dE_{\nu}},
\end{equation}
where $d_{\odot}$ is the Sun-Earth distance, $d\Phi_{\nu}/dE_{\nu}$
and $dN_{\nu}/dE_{\nu}$ represent the neutrino differential flux
at the Earth and the neutrino energy spectrum per DM annihilation
event in the Sun, respectively. The total annihilation rate $\Gamma_{A}$
can be expressed in terms of the number of the trapped DM particles
$N_{\chi}$:
\begin{eqnarray}
\varGamma_{A} & = & \frac{1}{2}A_{\odot}N_{\chi}^{2},
\end{eqnarray}
where $A_{\odot}$ denotes twice the annihilation rate of a pair of
DM particles. The evolution of the solar DM number $N_{\chi}$ is
depicted with the following equation:
\begin{equation}
\frac{dN_{\chi}}{dt}=C_{\odot}-E_{\odot}N_{\chi}-A_{\odot}N_{\chi}^{2},\label{eq:master equation-1}
\end{equation}
which involves the DM capture (evaporation) rate $C_{\odot}\,\left(E_{\odot}\right)$
by scattering off atomic nuclei in the Sun, as well as the annihilation
rate $A_{\odot}$. Eq.~(\ref{eq:master equation-1}) has an analytic
solution
\begin{equation}
N_{\chi}=\frac{C_{\odot}\,\tanh\left(t/\tau_{\mathrm{e}}\right)}{\tau_{\mathrm{e}}^{-1}+\left(E_{\odot}/2\right)\tanh\left(t/\tau_{\mathrm{e}}\right)},\label{eq:DMnumber}
\end{equation}
with

\begin{equation}
\tau_{\mathrm{e}}=\left(C_{\odot}\, A_{\odot}+E_{\odot}^{2}/4\right)^{-1/2}
\end{equation}
 the time scale for the capture, evaporation and annihilation processes
to equilibrate. Once the equilibrium is reached at the present day,
$i.e.$, $\tanh\left(t_{\odot}/\tau_{\mathrm{e}}\right)\simeq1$,
with $t_{\odot}=4.5\times10^{9}\,\mathrm{yr}$ being the solar age,
the annihilation output $\varGamma_{A}$ also reaches its maximum
value. As will be shown in Sec.~\ref{sub:evaporation}, a $\mathrm{GeV}$
increment in the DM mass parameter results in $1\sim2$ orders of
magnitude reduction in the evaporation rate $E_{\odot}$ in the few-$\mathrm{GeV}$
region. Thus depending on the ratio $E_{\odot}^{2}/\left(C_{\odot}A_{\odot}\right)$,
 such equilibrium can be categorized into
two different scenarios: (1) $E_{\odot}^{2}/\left(C_{\odot}A_{\odot}\right)\ll1$,
that's when the evaporation effect can be neglected and the equilibrium
is between annihilation and solar capture, $i.e.$, $\Gamma_{A}\simeq C_{\odot}/2$,
so we can either determine or constrain the strength of the DM-nucleon
interaction from solar neutrino observation; (2) $E_{\odot}^{2}/\left(C_{\odot}A_{\odot}\right)\gg1$,
under this circumstance evaporation overwhelms annihilation for the
DM depletion, and the balance between evaporation and solar capture
yields $\Gamma_{A}\simeq A_{\odot}C_{\odot}^{2}/\left(2\, E_{\odot}^{2}\right)$,
which not only implies a heavy suppression of the neutrino flux, but
also prevents us from drawing the coupling strength of the DM-nucleon
interaction from the possible observed signals.

Therefore, from the theoretical point of view it is interesting to
pin down the parameter space where the neutrino observation is relevant
for the DM detection. Conventionally, such purpose is fulfilled with
a characteristic quantity, the evaporation mass $m_{\mathrm{evap}}$,
which is defined with equation $E_{\odot}\left(m_{\mathrm{evap}}\right)=t_{\odot}^{-1}$
for the given DM-nucleon coupling. Above the evaporation mass one
can safely assume that the capture-annihilation equilibrium is reached.
The key point of the problem is to calculate the distribution of the
solar DM. While in Ref.~\cite{Spergel:1984re,Griest:1986yu} authors adopts a Maxwellian
distribution to describe the non-thermal equilibrium between the solar
DM particles and solar nuclei, the studies in Refs.~\cite{Nauenberg:1986em,Gould:1987ju}
indicate a deviation from the Maxwellian form, in a manner that the
actual velocity distribution is suppressed at the tail and tends to
be anisotropic at large radius. Such deviation can be attributed to
the fact that the energetic collisions that send the DM particles
into high orbits occur predominantly near the hot core of the Sun,
so as a result one expects a lower angular momentum distribution for
the high-energy orbits. In order to well describe the physical processes
such as evaporation and energy transfer of the solar DM, an accurate
description of the tail of the velocity distribution is necessary.

In addition, since the evaporation mass has been studied thoroughly
in the literature under the assumption of a constant DM-nucleon cross
section~\cite{Spergel:1984re,Nauenberg:1986em,Gould:1987ju,Busoni:2013kaa},
the quest to the extend the discussion to a broader set of DM-nucleon
effective interaction operators naturally arises. For instance, it
is tempting to evaluate the evaporation rate for the light asymmetric
DM particle with a DM-nucleon scattering amplitude linearly proportional
to the square of the transferred momentum $q^{2}$, because while
the authors of Refs.~\cite{Vincent:2014jia,Vincent:2015gqa} manage
to resolve the disagreement between the solar model and helioseismological
data with preferred DM mass of $3$ GeV and coupling strength of $10^{-37}\,\mathrm{cm}^{2}$,
the evaporation effect is not included in their discussion. Given
small DM masses as such, evaporation may no longer be neglected in
the buildup of the solar DM, and a quantitative analysis is needed
on this issue.

Thus, as a tentative study we investigate the implications of the
non-relativistic spin-independent (SI) effective operators on the
solar DM distribution and evaporation mass. The set of 15 Galilean
invariant operators is introduced in Ref.~\cite{Fitzpatrick:2012ix}\footnote[1]{\renewcommand{\baselinestretch}{1}\selectfont For an earlier important work on the non-relativistic effective theory of DM, see Ref.~\cite{Fan:2010gt}.}
as a comprehensive and convenient treatment for the DM-nucleus interaction
in the DM direct detection. Following Ref.~\cite{Gould:1987ju} we
calculate the non-thermal distribution of the solar DM by Monte Carlo
methods, and numerically compute evaporation rates for different SI
DM-nucleus effective operators. Moreover, based on the calculated
capture and evaporation rates, we also discuss the parameter space
relevant for the DM detection. This paper is organized as follows.
In Sec.~\ref{sec:EffectiveOperators} we take a brief review on the
effective interaction between the DM particle and nucleus. In Sec.~\ref{sec:Distribution=000026Evaporation}
we calculate the solar DM distribution and evaporation rate for various
SI DM-nucleus interaction operators, and discuss relevant implications
for the high-energy solar neutrino signals. Some interesting discussions
are arranged in Sec.~\ref{sec:Conclusion}.

\section{\label{sec:EffectiveOperators}Effective interaction between DM and
nucleus}

\begin{table}
\begin{centering}
\begin{tabular}{lcl}
 &  & \tabularnewline
\hline
\hline
$\hat{\mathcal{O}}_{1}=\mathbf{1}$ & $\qquad$ & $\hat{\mathcal{O}}_{9}=i\mathbf{\hat{\mathbf{S}}_{\chi}}\cdot\left(\hat{\mathbf{S}}_{N}\times\frac{\hat{\mathbf{q}}}{m_{N}}\right)$\tabularnewline
$\hat{\mathcal{O}}_{2}=\left(\hat{\mathbf{v}}^{\bot}\right)^{2}$ & \multirow{1}{*}{$\qquad$} & $\hat{\mathcal{O}}_{10}=i\left(\hat{\mathbf{S}}_{N}\cdot\frac{\hat{\mathbf{q}}}{m_{N}}\right)$\tabularnewline
$\hat{\mathcal{O}}_{3}=i\hat{\mathbf{S}}_{N}\cdot\left(\frac{\hat{\mathbf{q}}}{m_{N}}\times\hat{\mathbf{v}}^{\bot}\right)$ &  & $\hat{\mathcal{O}}_{11}=i\left(\mathbf{\hat{\mathbf{S}}_{\chi}}\cdot\frac{\hat{\mathbf{q}}}{m_{N}}\right)$\tabularnewline
$\hat{\mathcal{O}}_{4}=\mathbf{\hat{\mathbf{S}}_{\chi}}\cdot\hat{\mathbf{S}}_{N}$ & $\qquad$ & $\hat{\mathcal{O}}_{12}=\mathbf{\hat{\mathbf{S}}_{\chi}}\cdot\left(\hat{\mathbf{S}}_{N}\times\hat{\mathbf{v}}^{\bot}\right)$\tabularnewline
$\hat{\mathcal{O}}_{5}=i\mathbf{\hat{\mathbf{S}}_{\chi}}\cdot\left(\frac{\hat{\mathbf{q}}}{m_{N}}\times\hat{\mathbf{v}}^{\bot}\right)$ & $\qquad$ & $\hat{\mathcal{O}}_{13}=i\left(\mathbf{\hat{\mathbf{S}}_{\chi}}\cdot\hat{\mathbf{v}}^{\bot}\right)\left(\hat{\mathbf{S}}_{N}\cdot\frac{\hat{\mathbf{q}}}{m_{N}}\right)$\tabularnewline
$\hat{\mathcal{O}}_{6}=\left(\mathbf{\hat{\mathbf{S}}_{\chi}}\cdot\frac{\hat{\mathbf{q}}}{m_{N}}\right)\left(\hat{\mathbf{S}}_{N}\cdot\frac{\hat{\mathbf{q}}}{m_{N}}\right)$ & $\qquad$ & $\hat{\mathcal{O}}_{14}=i\left(\mathbf{\hat{\mathbf{S}}_{\chi}}\cdot\frac{\hat{\mathbf{q}}}{m_{N}}\right)\left(\hat{\mathbf{S}}_{N}\cdot\hat{\mathbf{v}}^{\bot}\right)$\tabularnewline
$\hat{\mathcal{O}}_{7}=\hat{\mathbf{S}}_{N}\cdot\hat{\mathbf{v}}^{\bot}$ & $\qquad$ & $\hat{\mathcal{O}}_{15}=-\left(\mathbf{\hat{\mathbf{S}}_{\chi}}\cdot\frac{\hat{\mathbf{q}}}{m_{N}}\right)\left[\left(\hat{\mathbf{S}}_{N}\times\hat{\mathbf{v}}^{\bot}\right)\cdot\frac{\hat{\mathbf{q}}}{m_{N}}\right]$\tabularnewline
\multicolumn{1}{l}{$\hat{\mathcal{O}}_{8}=\mathbf{\hat{\mathbf{S}}_{\chi}}\cdot\hat{\mathbf{v}}^{\bot}$} & \multicolumn{1}{c}{$\qquad$} & \tabularnewline
\hline
\hline
 &  & \tabularnewline
\end{tabular}
\par\end{centering}

\protect\caption{\label{tab:operators} A set of non-relativistic effective interaction
operators constructed from Eq.~(\ref{eq:buildingBlocks}) \cite{Fitzpatrick:2012ix}.
$m_{N}$ is the mass of the nucleon.}

\end{table}

We discuss the DM-nucleus scattering at low-energy scale in the context
of the non-relativistic (NR) effective interaction theory \cite{Fitzpatrick:2012ix,Anand:2013yka,Catena:2014uqa,Catena:2014hla,Catena:2014epa},
in which a set of linearly independent operators listed in Tab.~\ref{tab:operators}
can be generated from the following five Hermitian operators:
\begin{equation}
\mathbf{1},\quad i\hat{\mathbf{q}},\quad\hat{\mathbf{v}}^{\bot},\quad\hat{\mathbf{S}}_{\chi},\quad\hat{\mathbf{S}}_{N}.\label{eq:buildingBlocks}
\end{equation}
$\mathbf{q}$ is the transferred momentum from nucleon to the DM particle
in a collision, and the transverse velocity is defined as $\hat{\mathbf{v}}^{\bot}=\mathbf{v}+\mathbf{q}/\left(2\mu_{N}\right)$,
which satisfies $\mathbf{q}\cdot\hat{\mathbf{v}}^{\bot}=0$ for the
on-shell process, where $\mathbf{v}=\mathbf{v}_{\chi,i}-\mathbf{v}_{N,i}$
is the relative initial velocity between the DM particle and nucleon,
and $\mu_{N}=m_{\chi}m_{N}/\left(m_{\chi}+m_{N}\right)$ is the reduced
mass of the system. $\hat{\mathbf{S}}_{\chi}$ and $\hat{\mathbf{S}}_{N}$
are the spins of the DM particle and the nucleon, respectively.

While the operators presented in Tab.~\ref{tab:operators} exhaust
all the possible NR reduction of the Lorentz invariant spin-1/2 DM-nucleon
interaction, up to corresponding coefficients dependent on the Galilean
invariant scalar $q^{2}$, in this study we shall investigate the
implication of all the SI operators $\hat{\mathcal{O}}_{1}$, $\hat{\mathcal{O}}_{5}$,
$\hat{\mathcal{O}}_{8}$ and $\hat{\mathcal{O}}_{11}$\footnote[2]{\renewcommand{\baselinestretch}{1}\selectfont $\hat{\mathcal{O}}_{2}$ is out of consideration
because it will not be induced as the leading order term in non-relativistic
expansion from the relativistic operators, unless there exists  significant fine tuning that leads to a
delicate cancellation among the leading pieces~\cite{Fitzpatrick:2012ix}.} for the DM
evaporation mass.

Since the atomic nucleus is a composite of bound nucleons, its structural
effect has to be taken into consideration in the analysis of the DM-nucleus
interaction. Interestingly, in addition to the conventional nuclear
form factor that describes the mass distribution within a nucleus,
other types of DM and nuclear response functions arise from various
underlying DM-nucleon interactions. For example, the operator $\hat{\mathbf{v}}^{\bot}$
can be divided into the centre-of-mass and the relative motion components
as
\begin{eqnarray}
\hat{\mathbf{v}}^{\bot} & = & \hat{\mathbf{v}}_{\mathcal{A}}^{\bot}-\frac{1}{2}\left(\mathbf{v}_{N,i}-\mathbf{v}_{\mathcal{A},i}+\mathbf{v}_{N,f}-\mathbf{v}_{\mathcal{A},f}\right)\nonumber \\
 & = & \mathbf{v}_{\chi,i}-\mathbf{v}_{\mathcal{A},i}+\frac{\mathbf{q}}{2\mu_{\mathcal{A}}}-\frac{1}{2}\left(\mathbf{v}_{N,i}-\mathbf{v}_{\mathcal{A},i}+\mathbf{v}_{N,f}-\mathbf{v}_{\mathcal{A},f}\right),
\end{eqnarray}
where $\mu_{\mathcal{A}}$ is reduced mass of the DM-nucleus system,
$\mathbf{v}_{N,i}\,\left(\mathbf{v}_{N,f}\right)$ and $\mathbf{v}_{\mathcal{A},i}\,\left(\mathbf{v}_{\mathcal{A},f}\right)$
denote the initial (final) velocities of the constituent nucleon and
the whole nucleus, respectively. While $\hat{\mathbf{v}}_{\mathcal{A}}^{\bot}\equiv\mathbf{v}_{\chi,i}-\mathbf{v}_{\mathcal{A},i}+\mathbf{q}/\left(2\mu_{\mathcal{A}}\right)$
represents the nucleus transverse velocity, the latter term $\frac{1}{2}\left(\mathbf{v}_{N,i}-\mathbf{v}_{\mathcal{A},i}+\mathbf{v}_{N,f}-\mathbf{v}_{\mathcal{A},f}\right)$
corresponds to the convection current operator $\frac{1}{2m_{N}}\left(i\overleftarrow{\nabla}_{x_{N}}\delta^{3}\left(\mathbf{x}_{N}-\mathbf{x}_{\mathcal{A}}\right)+\delta^{3}\left(\mathbf{x}_{N}-\mathbf{x}_{\mathcal{A}}\right)\left(-i\right)\overrightarrow{\nabla}_{x_{N}}\right)$
in  coordinate space, and gives rise to a nuclear response function
($\Delta$ response in Ref.~\cite{Fitzpatrick:2012ix}) associated
with the nuclear orbital angular momentum in the long-wavelength limit. Nevertheless,
compared with the conventional form factor that corresponds to $W_{M}$
in Refs.~\cite{Anand:2013yka,Catena:2014uqa,Catena:2014hla,Catena:2014epa},
response functions coming from the nuclear intrinsic motion ($W_{\Delta}$
in Refs.~\cite{Anand:2013yka,Catena:2014uqa,Catena:2014hla,Catena:2014epa})
can be safely neglected if the isospin symmetry is respected. This
is a direct observation from the nuclear response functions provided
in Ref.~\cite{Catena:2015uha}: isoscalar response functions $\left(\mu_{\mathcal{A}}/m_{N}\right)^{2}W_{\Delta}^{00}$
are much smaller than $W_{M}^{00}$ for the unpaired solar
elements ($e.g.$, $^{14}\mathrm{N}$, $^{23}\mathrm{Na}$ and $^{27}\mathrm{Al}$).
Not even to mention that these $W_{\Delta}$ responses associated
with the unpaired elements suffer significant abundance suppression
in the Sun.

\begin{table}
\begin{centering}
\begin{tabular}{llc}
 &  & \tabularnewline
\hline
\hline
$\hat{\mathcal{O}}_{i}$ & $\qquad$ & \multirow{1}{*}{$P_{i}\left(v_{\mathrm{rel}}^{2},\, q^{2}\right)$}\tabularnewline
\hline
$\hat{\mathcal{O}}_{1}$ & $\qquad$ & 1\tabularnewline
$\hat{\mathcal{O}}_{5}$ & $\qquad$ & $\frac{j_{\chi}\left(j_{\chi}+1\right)}{3}\frac{q^{2}}{m_{N}^{2}}v_{\mathcal{A}}^{\bot2}$\tabularnewline
$\hat{\mathcal{O}}_{8}$ & $\qquad$ & $\frac{j_{\chi}\left(j_{\chi}+1\right)}{3}v_{\mathcal{A}}^{\bot2}$\tabularnewline
\multicolumn{1}{l}{$\hat{\mathcal{O}}_{11}$} & \multicolumn{1}{l}{$\qquad$} & $\frac{j_{\chi}\left(j_{\chi}+1\right)}{3}\frac{q^{2}}{m_{N}^{2}}$\tabularnewline
\hline
\hline
 &  & \tabularnewline
\end{tabular}
\par\end{centering}

\protect\caption{\label{tab:DM response }The DM response functions for operators $i=1,\,5,\,8,\,\mathrm{and}\,11$.
See text for details. }
\end{table}

Therefore, assuming the DM particle couples to the proton and neutron
with equal strengths, the effects of response $\Delta$ can be neglected
for operators $\hat{\mathcal{O}}_{5}$ and $\hat{\mathcal{O}}_{8}$,
and hence we simply utilize the conventional Helm form factor to account
for the nuclear internal structure, when investigating the implication
of various SI interactions on the DM evaporation on a case-by-case
basis. As a result,  the DM-nucleus differential cross section for
operators $i=1,\,5,\,8,\,11$ can be expressed in terms of the transferred
momentum $q$ as follows
\begin{eqnarray}
\frac{d\sigma_{i}}{dq} & = & \frac{c_{i}^{2}A^{2}F_{N}^{2}\left(q^{2}\right)}{2\pi v_{\mathrm{rel}}^{2}}P_{i}\left(v_{\mathrm{rel}}^{2},\, q^{2}\right)q,\label{eq:cross section}
\end{eqnarray}
where $c_{i}$ carrying a dimension of $\mathrm{mass^{-2}}$ is the
nucleon coupling constant for operator $\hat{\mathcal{O}}_{i}$, $A$
is the atomic number of the target nucleus $\mathcal{A}$, $\mathbf{v}_{\mathrm{rel}}=\mathbf{v}_{\chi,i}-\mathbf{v}_{\mathcal{A},i}$
is the relative incoming velocity of the DM-nucleus system, and $P_{i}\left(v_{\mathrm{rel}}^{2},\, q^{2}\right)$
is the corresponding DM response function listed explicitly in Tab.~\ref{tab:DM response }.
In Tab.~\ref{tab:DM response }, $j_{\chi}$ represents the spin
of the DM particle,  and $v_{\mathcal{A}}^{\bot2}=v_{\mathrm{rel}}^{2}-q^{2}/\left(4\mu_{\mathcal{A}}^{2}\right)$
when the on-shell requirement is satisfied.  $F_{N}^{2}(q^{2})=\left[3\,\mathrm{j_{1}}\left(q\, R_{1}\right)/\left(q\, R_{1}\right)\right]^{2}e^{-q^{2}s^{2}}$
is the Helm form factor, with $\mathrm{j_{1}}(x)=\sin\left(x\right)/x^{2}-\cos\left(x\right)/x$
being the Bessel spherical function of the first kind, $R_{1}=\sqrt{R_{0}^{2}-5s^{2}}$
with $R_{0}\backsimeq1.23\, A^{1/3}\,\mathrm{fm}$, and $s\backsimeq1\,\mathrm{fm}$
\cite{Lewin:1995rx}.

\section{\label{sec:Distribution=000026Evaporation}Distribution and Evaporation
of solar DM}

In this section we will discuss the distribution and evaporation of
the solar DM. Since the evaporation occurs predominantly at the high
end of the velocity distribution, its evaluation relies on an accurate
description thereof. We determine the solar DM distribution by solving
the Boltzmann equation in a numerical way, and then separately calculate
the evaporation rate for various effective SI DM-nucleon interaction
operators. Now we delve into the details.

\subsection{\label{sub:sec.3.a}high end of the velocity distribution in the
Sun}

To date, there are two effective strategies in literature for determining
the solar DM distribution. In the ``Brownian motion'' method that
is pioneered by the author of Ref.~\cite{Nauenberg:1986em}, the
distribution sample is obtained by simulating the motion of a single
DM particle wandering in the Sun \footnote[3]{\renewcommand{\baselinestretch}{3}\selectfont See Appendix A in Ref.~\cite{Chen:2015uha} for an example.}. While the ``Brownian motion''
method is efficient in describing the bulk of the velocity distribution,
it turns impractical in computing the tail of the distribution for
which a huge and uneconomical base of event samples is required to
generate sufficient statistics. Therefore in order to determine the
distribution of the solar DM, we resort to essentially the same method
as the one outlined in Ref.~\cite{Gould:1987ju}.

Here we take a brief introduction to the methodology. Our discussion
begins with the assumption that the presence of the solar DM does
$not$ bring any significant impact on the solar structure, $i.e.$,
the feedback from the accumulating DM particles is assumed to be negligible.
The Boltzmann equation is linear due to the absence of the DM self-interaction,
and can be further simplified as the following master equation if
expressed with a convenient choice of parameters $E$ (total energy per unit mass)
and $L$ (angular momentum per unit mass) \cite{Gould:1987ju}:
\begin{eqnarray}
\frac{\mathrm{d}f\left(E,\, L\right)}{\mathrm{d}t} & = & -f\left(E,\, L\right)\sum_{E',L'}S\left(E,\, L;\, E',\, L'\right)+\sum_{E',L'}f\left(E',\, L'\right)S\left(E',\, L';\, E,\, L\right),\label{eq:master equation}
\end{eqnarray}
where $f\left(E,\, L\right)$ is the distribution function of the
solar DM, and $S\left(E,\, L;\, E',\, L'\right)$ represents the scattering
matrix element for transition process $\left(E,\, L\right)\rightarrow\left(E',\, L'\right)$.
In fact, to fully describe the physical state of the bound DM particle
we still need an extra parameter, say, a temporal parameter $\tau$,
to label the position in the periodic orbit defined by energy and
angular momentum. However, we approximate both the distribution function
and scattering matrix elements as independent of parameter $\tau$
in Eq.~(\ref{eq:master equation}). The reason is because a small
DM-nucleus cross section, or equivalently, a large mean free path
leads to a slowly increasing probability for a renewal collision,
which implies an insensitive reliance of the distribution and scattering
matrix on parameter $\tau$.

The scattering matrix $S\left(E,\, L;\, E',\, L'\right)$ is determined
with simulation approach and the $weighting$ method is adopted to
facilitate the computation. Specifically speaking, we first calculate
the probability for a trapped DM particle to collide with the solar
elements on its trajectory at a fixed time interval $\Delta t$, and
then as a weight this probability is multiplied with the tally of
the simulating transition events, so as to evaluate the scattering
matrix in a more efficient manner. The numerical integration of the
bound DM orbits is based on the Standard Sun Model (SSM) GS98 \cite{Serenelli:2009yc} and 5 solar elements $\mathrm{H}$, $\mathrm{^{4}He}$, $^{14}\mathrm{N}$,
$^{16}\mathrm{O}$ and $^{56}\mathrm{Fe}$ are included in the simulation
of the DM-nucleus scattering. With random numbers that help pick out both the colliding
solar element and its velocity, as well as the scattering angle in the centre-of-mass (CM) frame, we
determine the outgoing state of the scattered DM particle after a
coordinate transformation back to the solar reference. Further details
of the discussion on the thermal collision are arranged in Appendix~\ref{sec:Appendixa}.
\begin{figure}
\begin{centering}
\includegraphics[scale=0.7]{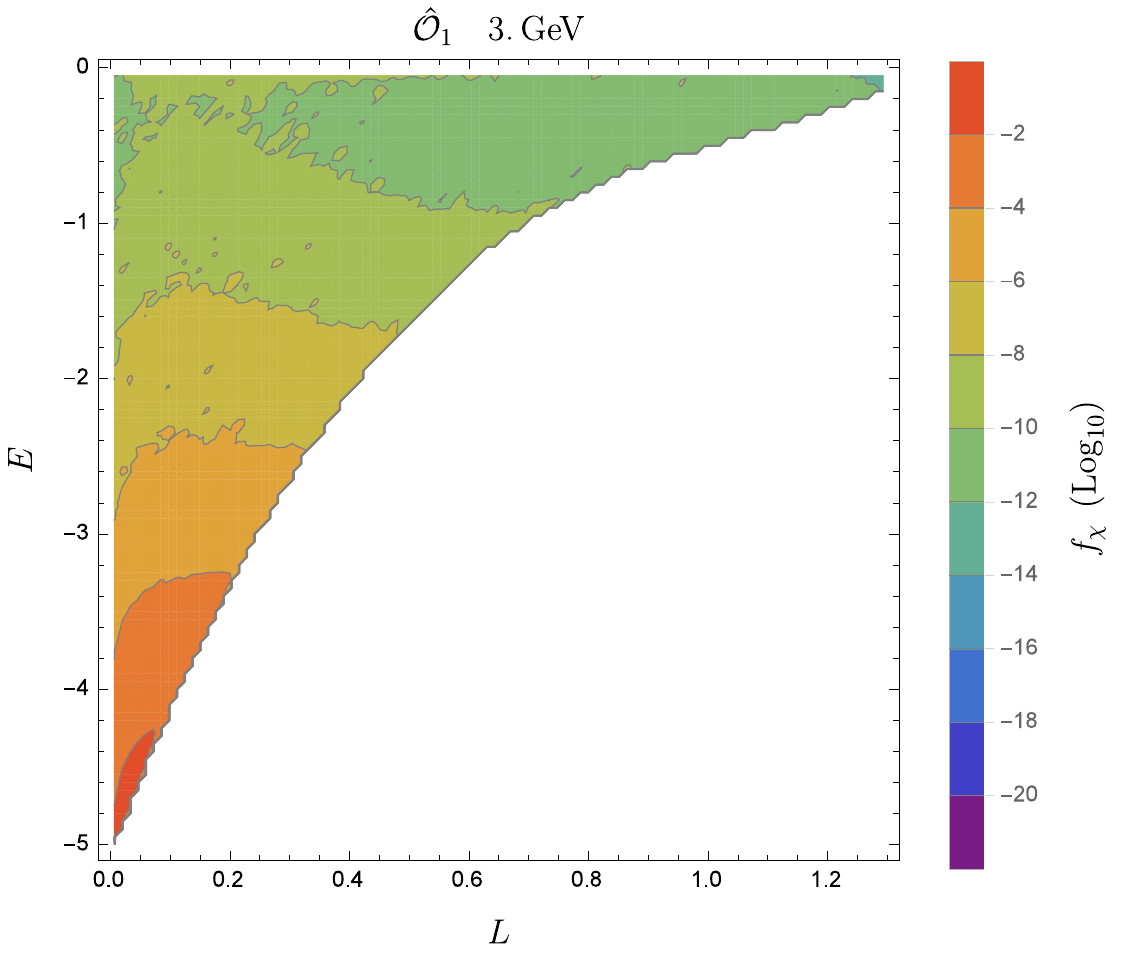} \includegraphics[scale=0.7]{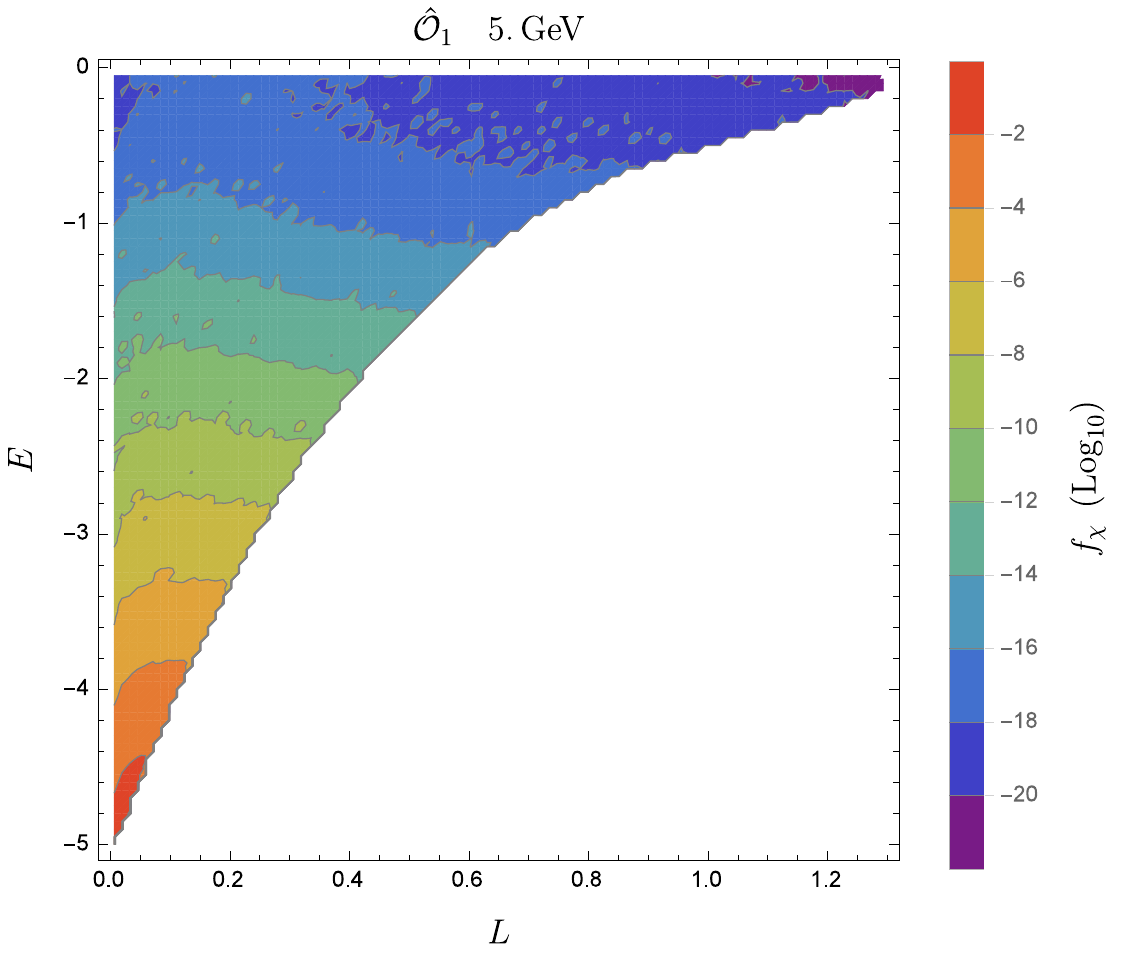}
\protect\caption{\label{fig:fel}The equilibrium distribution $f_{\chi}\left(E,\, L\right)$
for operator $\hat{\mathcal{O}}_{1}$ at $m_{\chi}=3\,\mathrm{GeV}$
($left$) and $m_{\chi}=5\,\mathrm{GeV}$ ($right$). The energy $E$
and angular momentum $L$ are nondimensionalised in units of $GM_{\odot}/R_{\odot}$
and $\sqrt{GM_{\odot}R_{\odot}}$, respectively. Only the coloured
parameter region is allowed for bound orbits. See text for details.}

\par\end{centering}

\end{figure}

It is also worth  mentioning that in principle all kinetically allowed states of $\left(E,\, L\right)$, including both the bound and unbound states that are connected to each other through  capture and evaporation, should be involved in Eq.~(\ref{eq:master equation}) for  a realistic description of the solar DM.
In practice, however, we model the  captured DM particles as a closed system; that
is to say,  the number of the  solar DM particles  is  assumed to be conserved within a timescale comparable to the relaxation time of the system, and the  transitions are confined to only the gravitational bound states.
The validity of this assumption will be discussed in Sec.~\ref{sec:Conclusion}. As a consequence,
 Eq.~(\ref{eq:master equation})
represents a Markov process. We evolve it with the discrete time step
$\Delta t$ until $f\left(E,\, L\right)$ converges to the limiting
distribution $f_{\chi}\left(E,\, L\right)$. For illustration, we
present the equilibrium distribution $f_{\chi}\left(E,\, L\right)$
for the DM-nucleon interaction operator $\hat{\mathcal{O}}_{1}$ in
Fig.~\ref{fig:fel}. The parameters $E$ and $L$ are nondimensionalised
in units of an energy reference value $GM_{\odot}/R_{\odot}$, and
an angular momentum value $\left(GM_{\odot}R_{\odot}\right)^{1/2}$, where $G$ is the Newton's constant, and $M_{\odot}$ is the solar mass. These values are constructed from a length unit, namely the solar
radius $R_{\odot}=6.955\times10^{5}\,\mathrm{km}$, and a time unit
$\left(GM_{\odot}/R_{\odot}^{3}\right)^{-1/2}=1.596\times10^{3}\,\mathrm{s}$,
from which the DM velocity $v_{\chi}$ can also be expressed in terms
of a  reference value $\left(GM_{\odot}/R_{\odot}\right)^{1/2}\approx436\,\mathrm{km\cdot s^{-1}}$.

\begin{figure}
\begin{centering}
\includegraphics[scale=0.6]{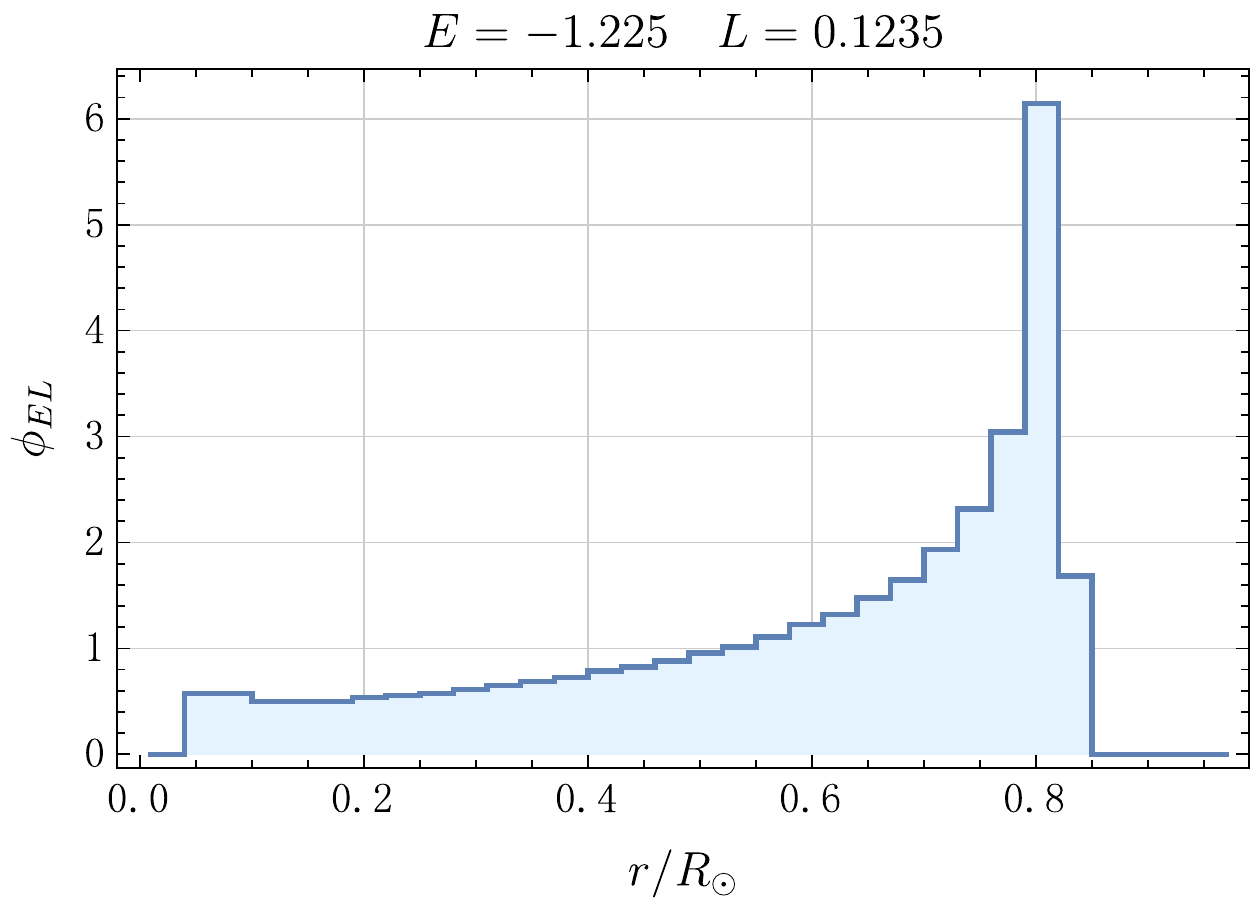}$\quad$\includegraphics[scale=0.6]{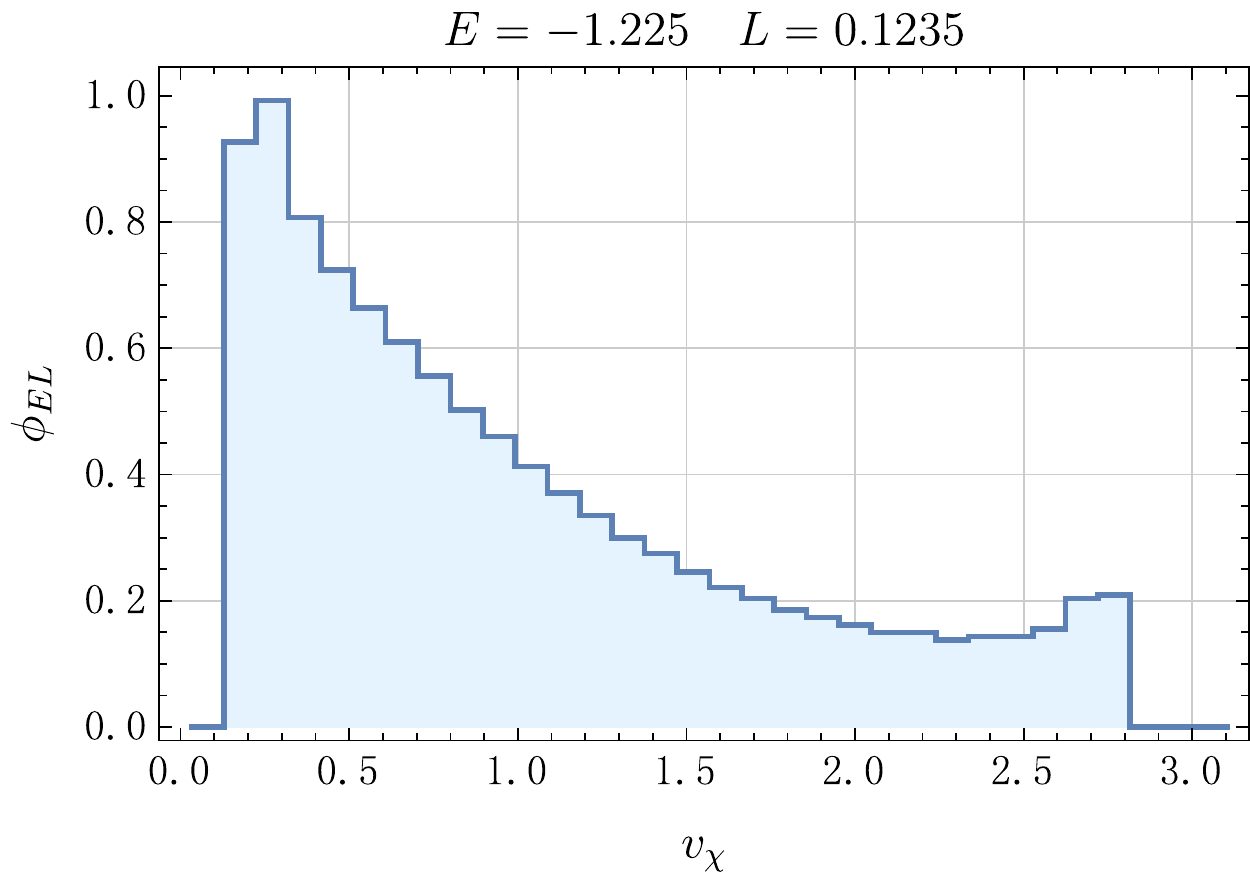}
\par\end{centering}

\protect\caption{\label{fig:Phi}The integrated distribution function of radius $r$
$\left(left\right)$ and velocity $v_{\chi}$ $\left(right\right)$
respectively for the orbit $\left(E=-1.225,\, L=0.124\right)$. }
\end{figure}

Finally, by convoluting $f_{\chi}\left(E,\, L\right)$ with $\phi_{EL}\left(\, r,\, v_{\chi}\right)$,
the distribution function of radius $r$ and velocity $v_{\chi}$
for orbit $\left(E,\, L\right),$ we obtain the DM distribution function
\begin{eqnarray}
f_{\chi}\left(r,\, v_{\chi}\right) & = & \sum_{E,L}f_{\chi}\left(E,\, L\right)\phi_{EL}\left(\, r,\, v_{\chi}\right).
\end{eqnarray}
For illustration, we present the distribution function of radius $r$
after integrating out velocity $v_{\chi}$ and $vice\, versa$ for
the orbit $E=-1.225,\, L=0.124$ in Fig.~\ref{fig:Phi}.

\begin{figure}
\begin{centering}
\includegraphics[scale=0.57]{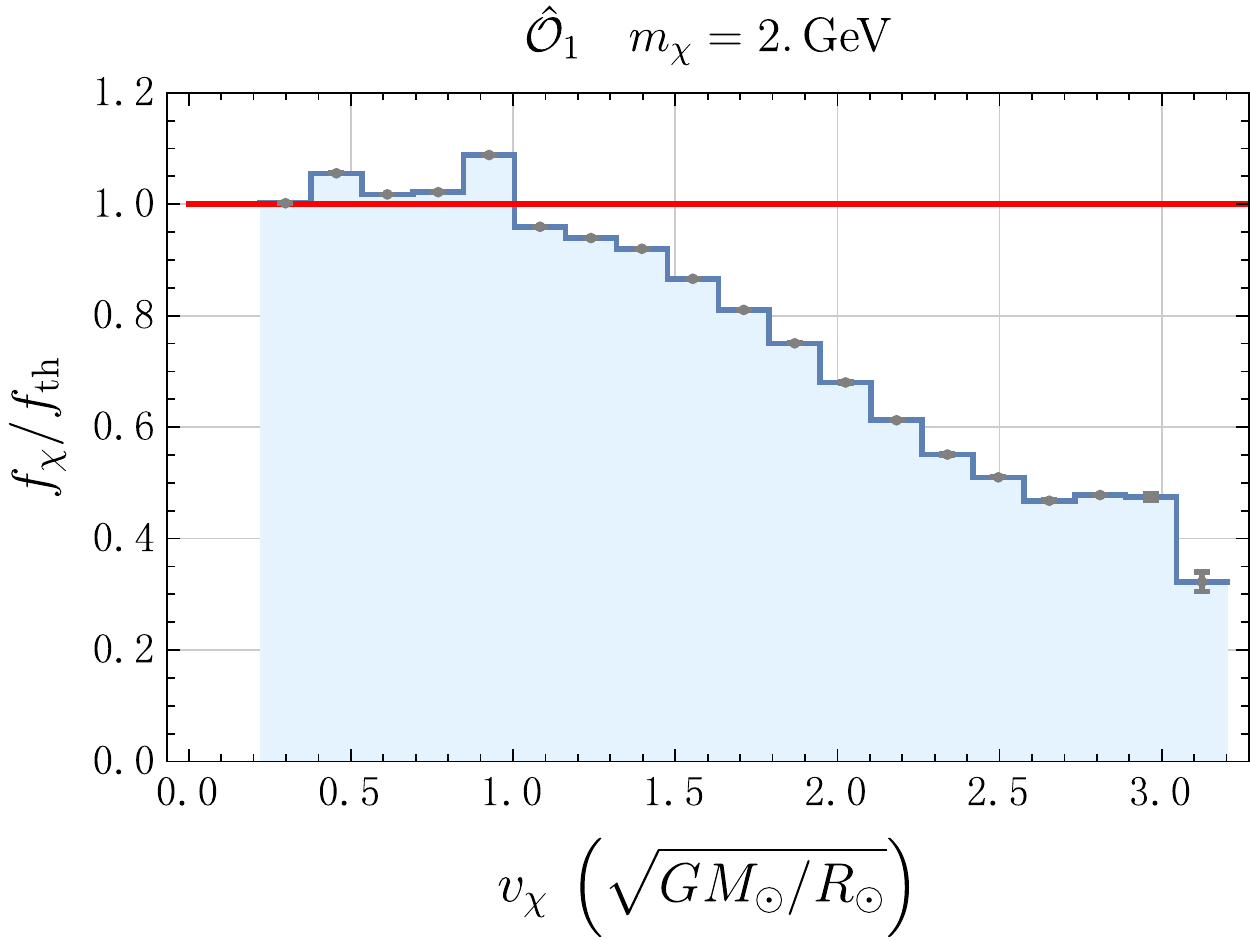}$\quad$\includegraphics[scale=0.57]{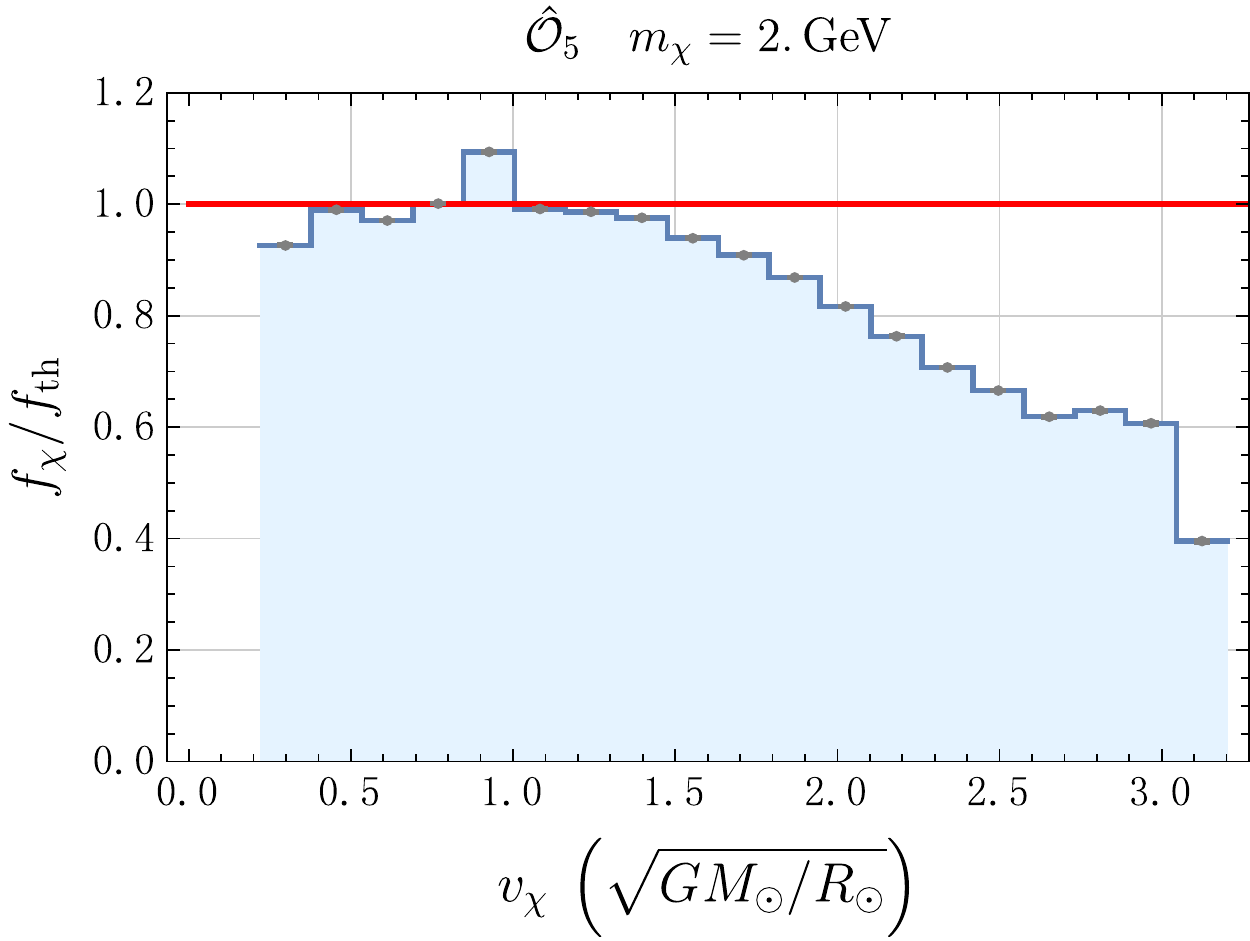}\vspace{0.1cm}

\par\end{centering}

\begin{centering}
\includegraphics[scale=0.57]{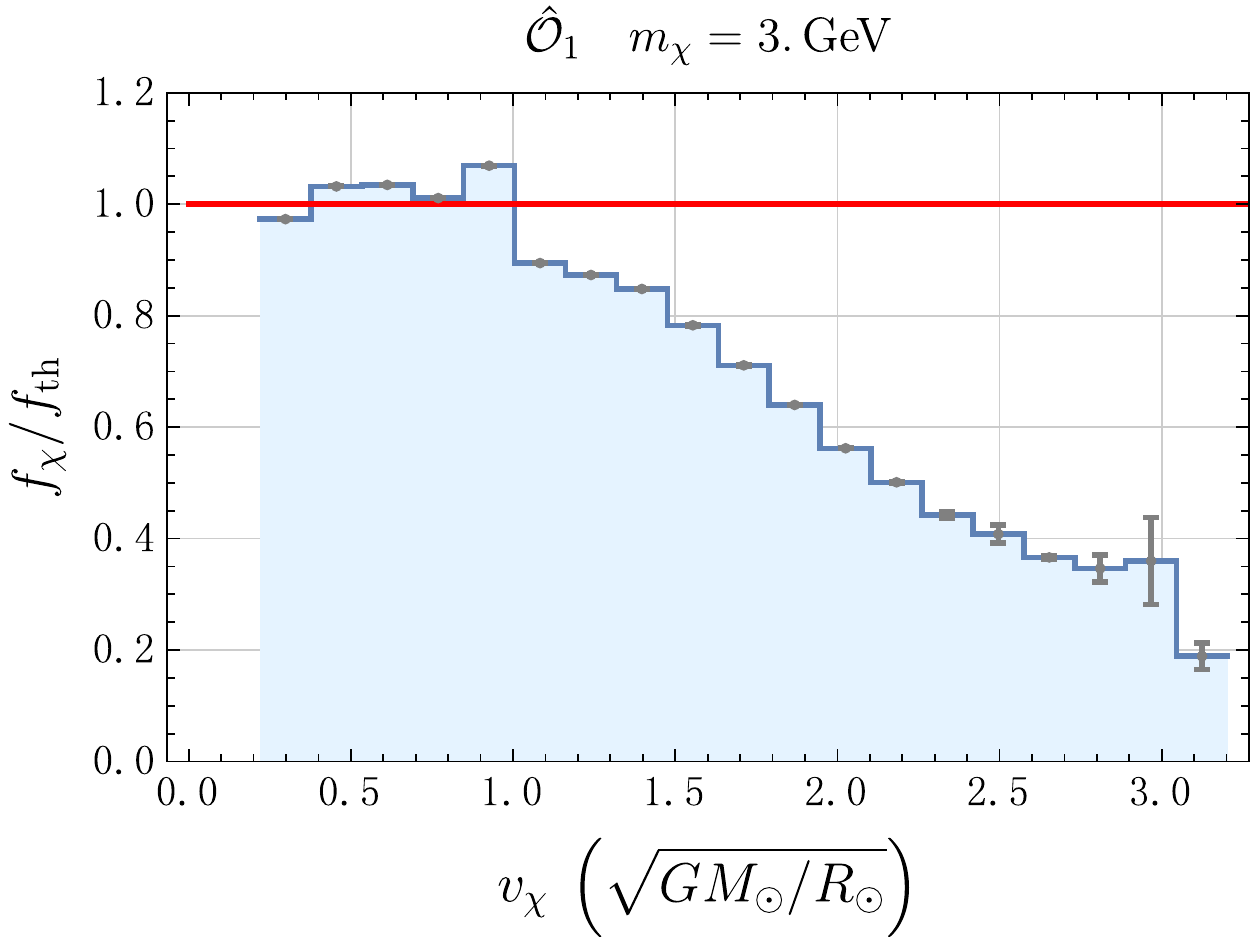}$\quad$\includegraphics[scale=0.57]{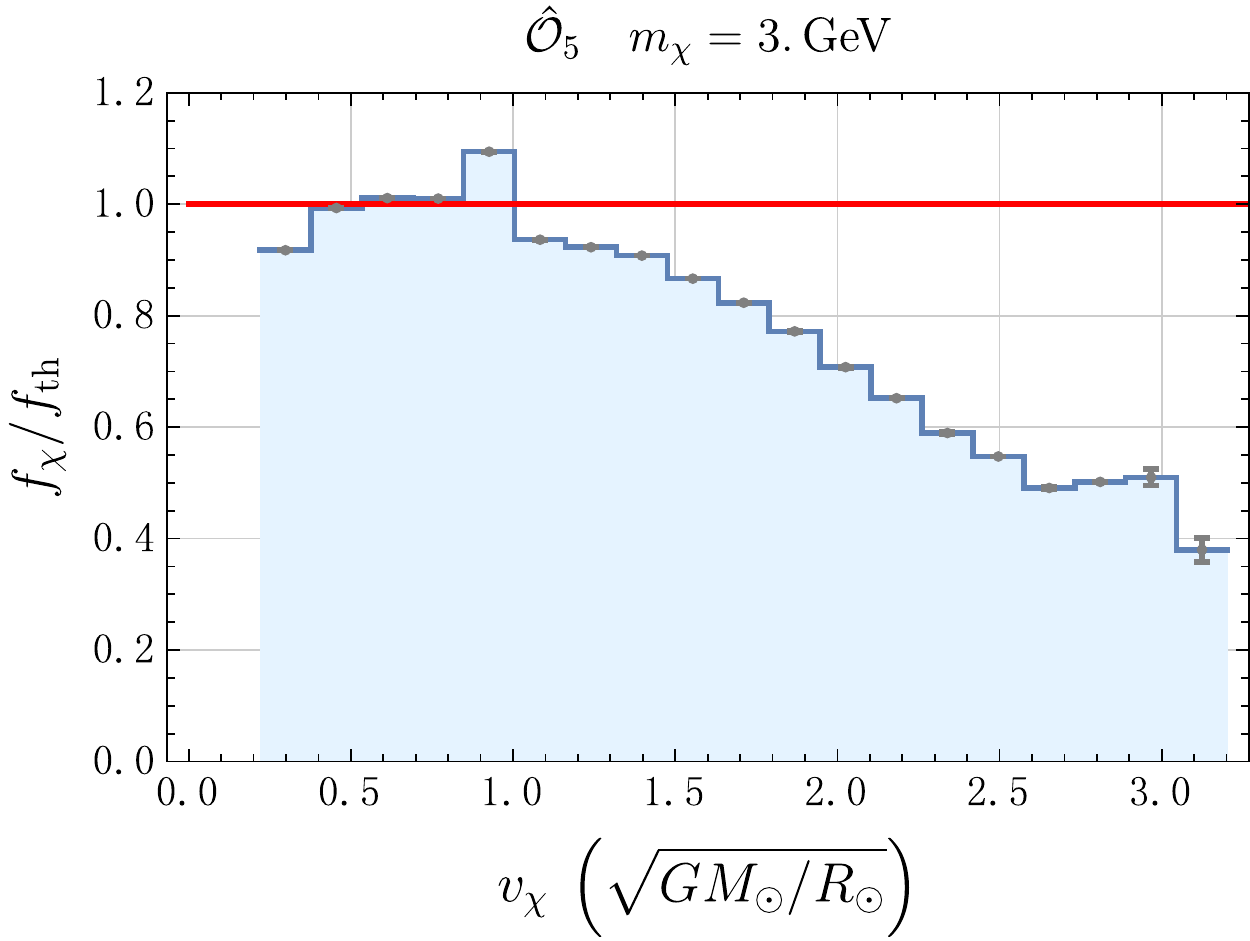}\vspace{0.1cm}

\par\end{centering}

\begin{centering}
\includegraphics[scale=0.57]{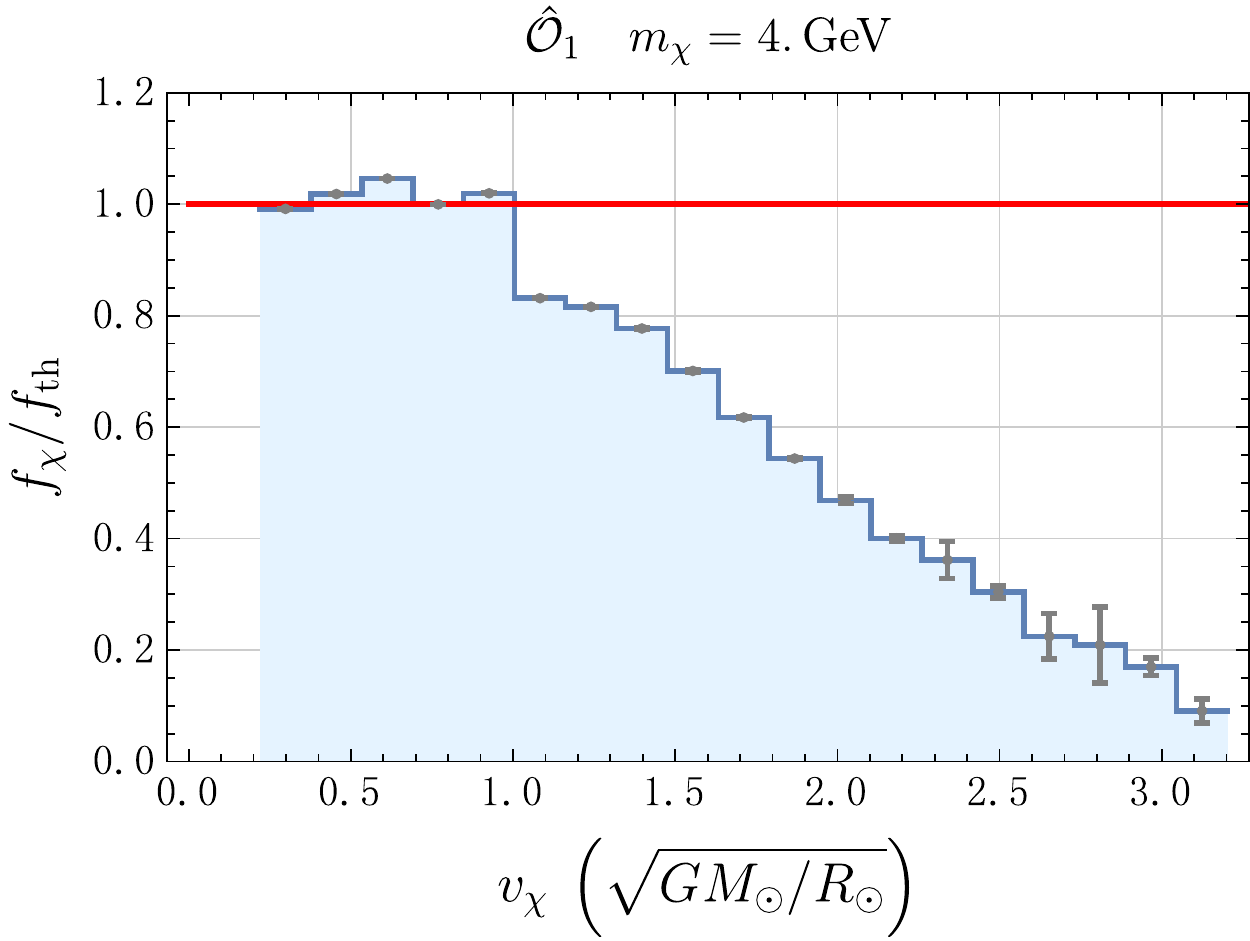}$\quad$\includegraphics[scale=0.57]{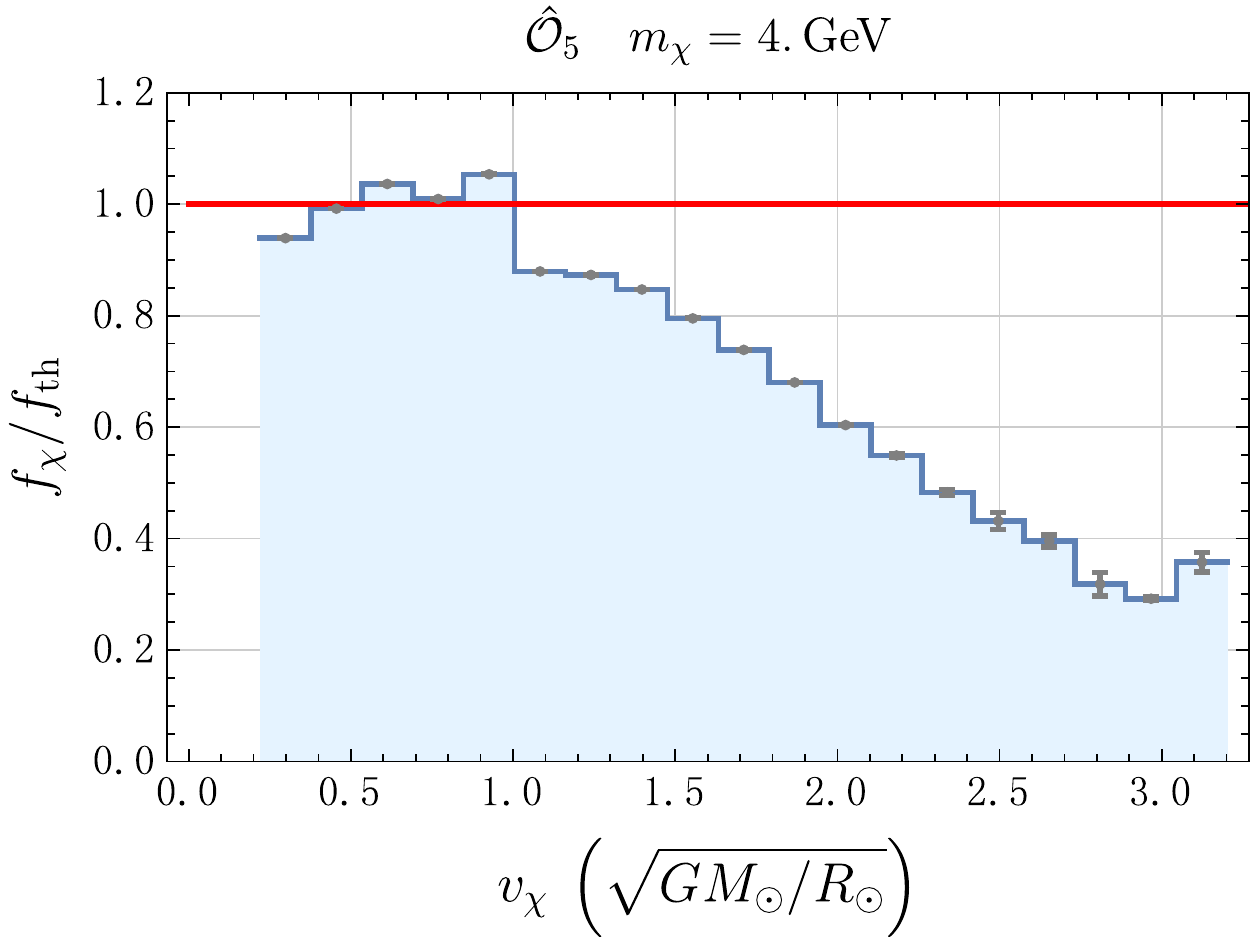}
\par\end{centering}

\begin{centering}
\vspace{0.1cm}

\par\end{centering}

\begin{centering}
\includegraphics[scale=0.57]{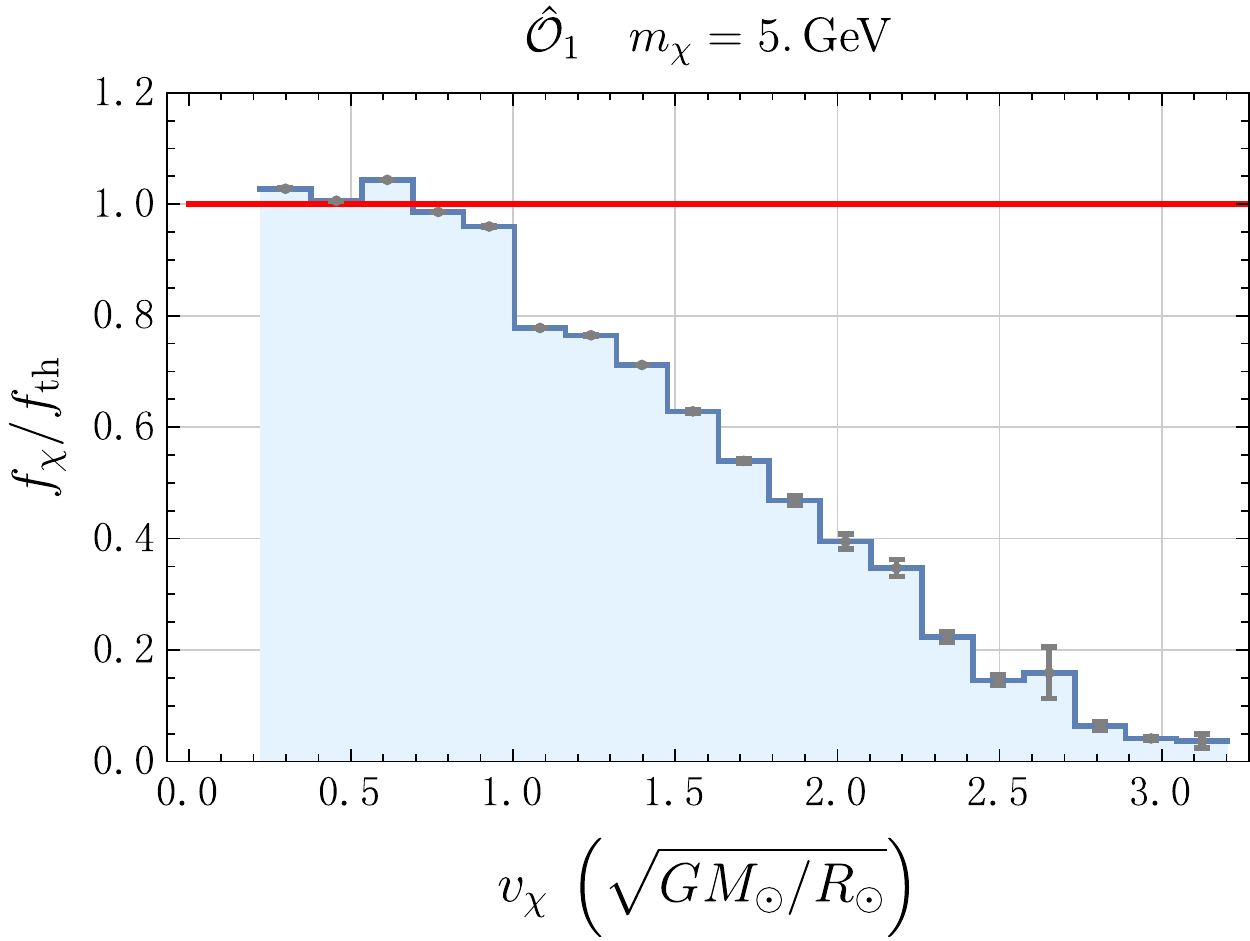}$\quad$\includegraphics[scale=0.57]{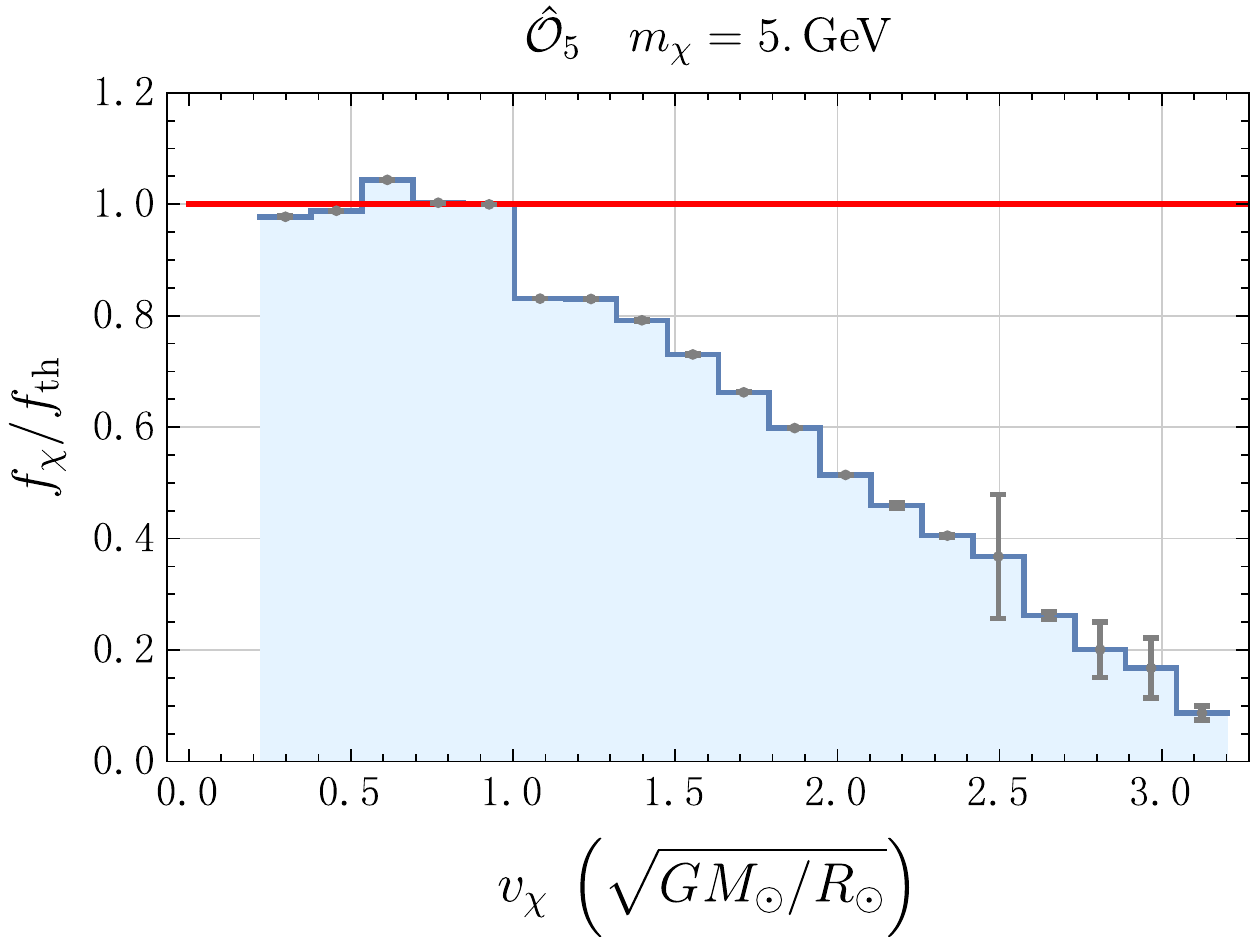}
\par\end{centering}

\protect\caption{\label{fig:RatioVelocity1}The ratio of the non-thermal distribution
to the approximated thermal distribution as a function of DM mass
at $m_{\chi}=2.0$, $3.0$, $4.0$ and $5.0$ GeV, for effective operators
$\hat{\mathcal{O}}_{1}$ ($left$) and $\hat{\mathcal{O}}_{5}$ ($right$),
respectively. }
\end{figure}

\begin{figure}
\begin{centering}
\includegraphics[scale=0.57]{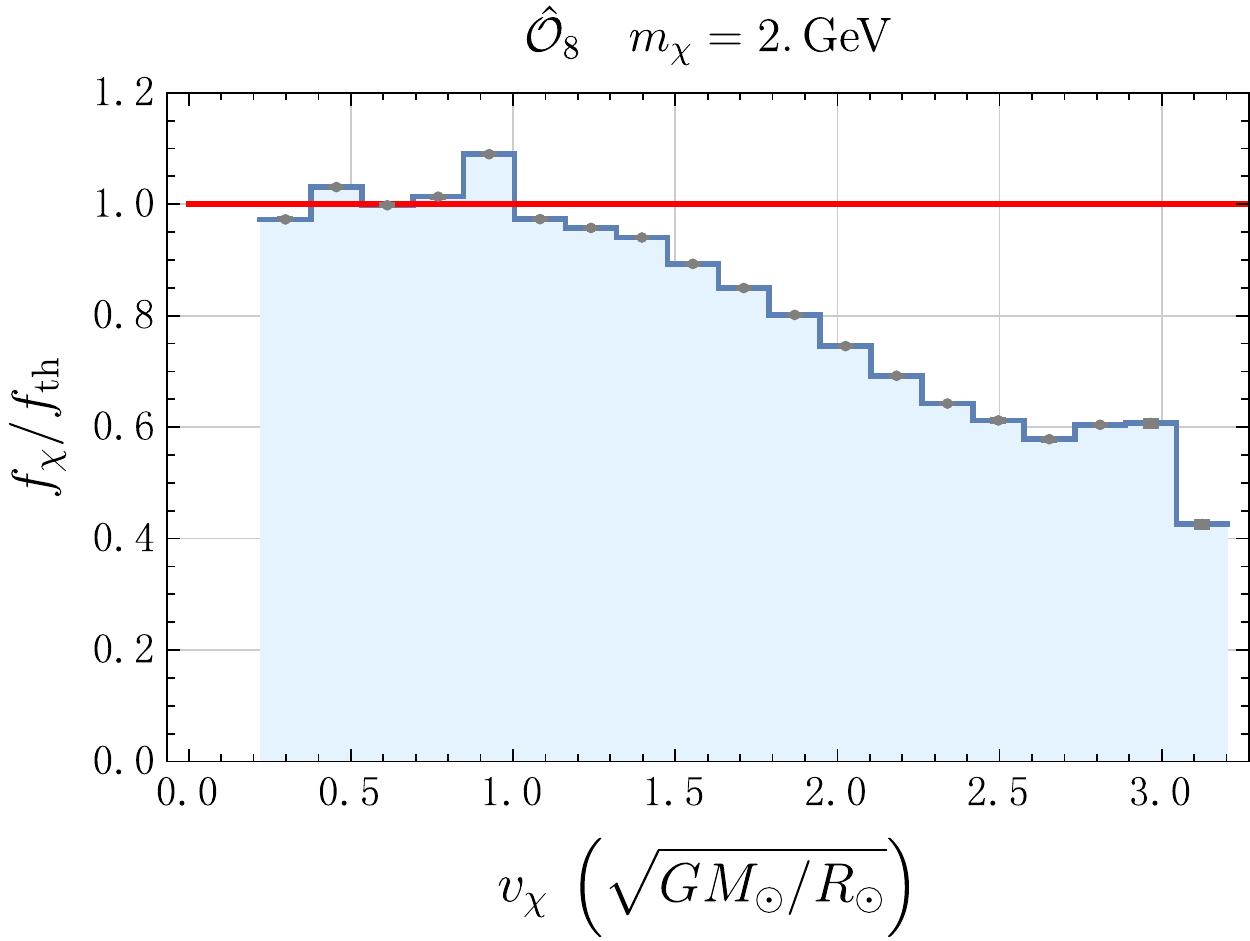}$\quad$\includegraphics[scale=0.57]{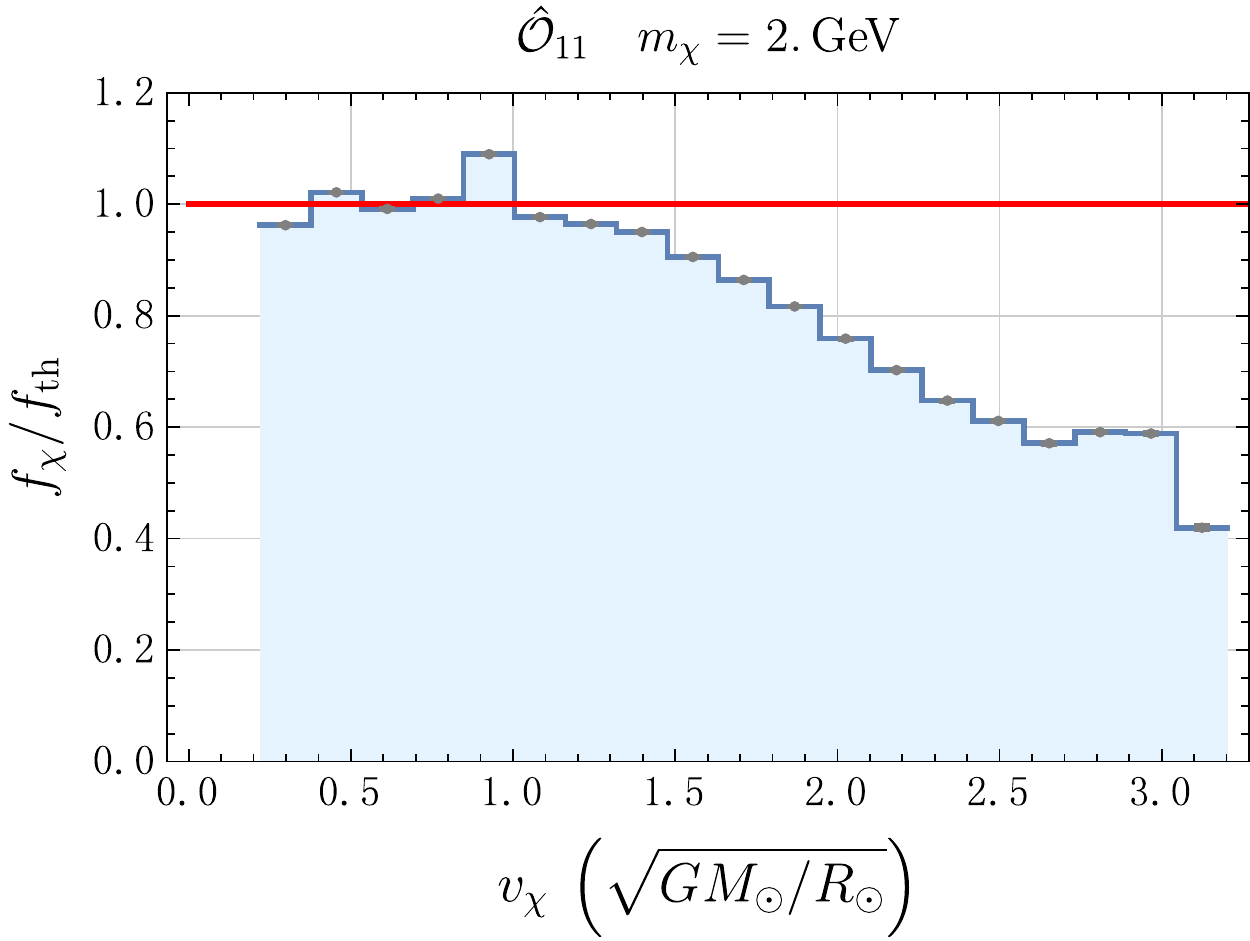}\vspace{0.1cm}

\par\end{centering}

\begin{centering}
\includegraphics[scale=0.57]{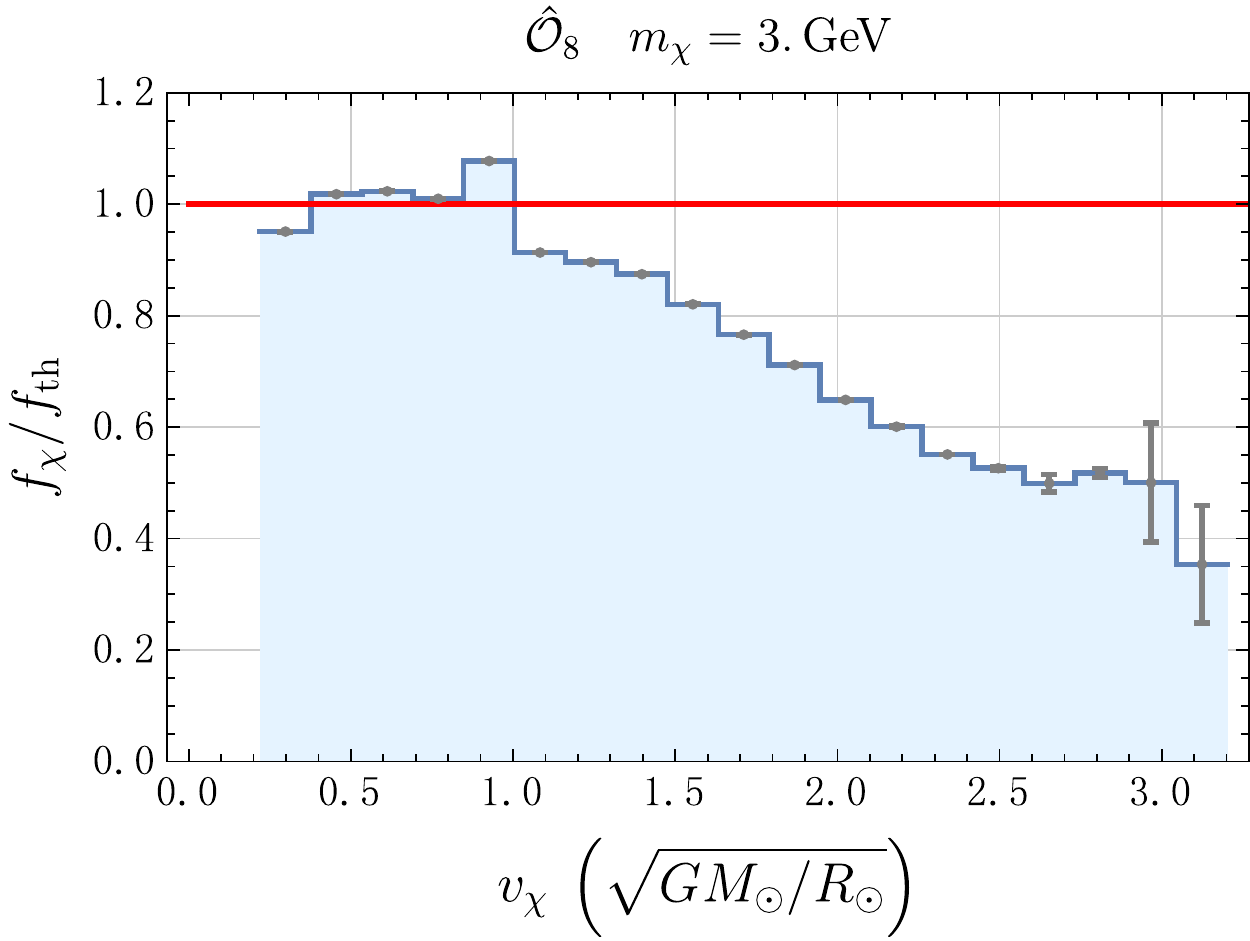}$\quad$\includegraphics[scale=0.57]{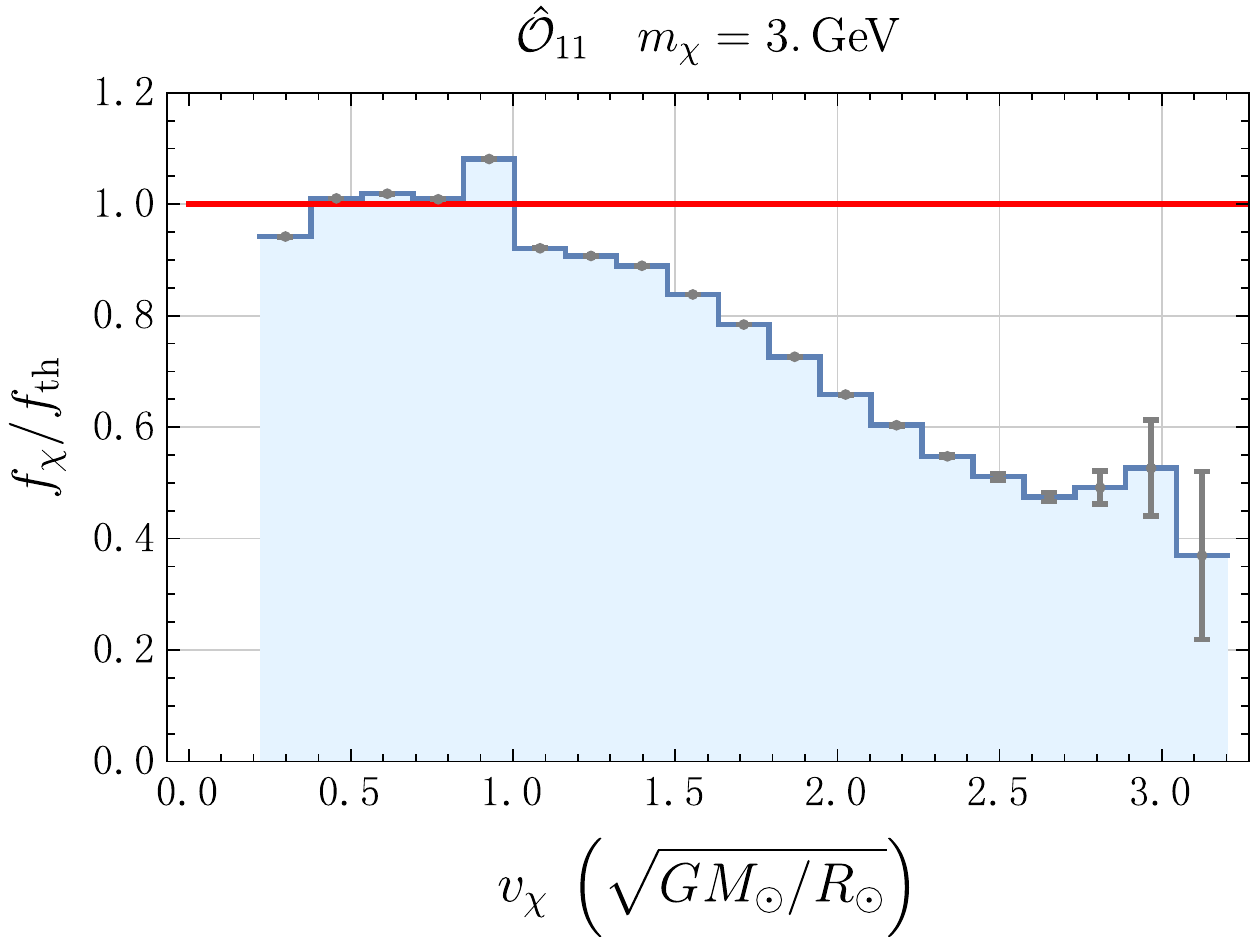}\vspace{0.1cm}

\par\end{centering}

\begin{centering}
\includegraphics[scale=0.57]{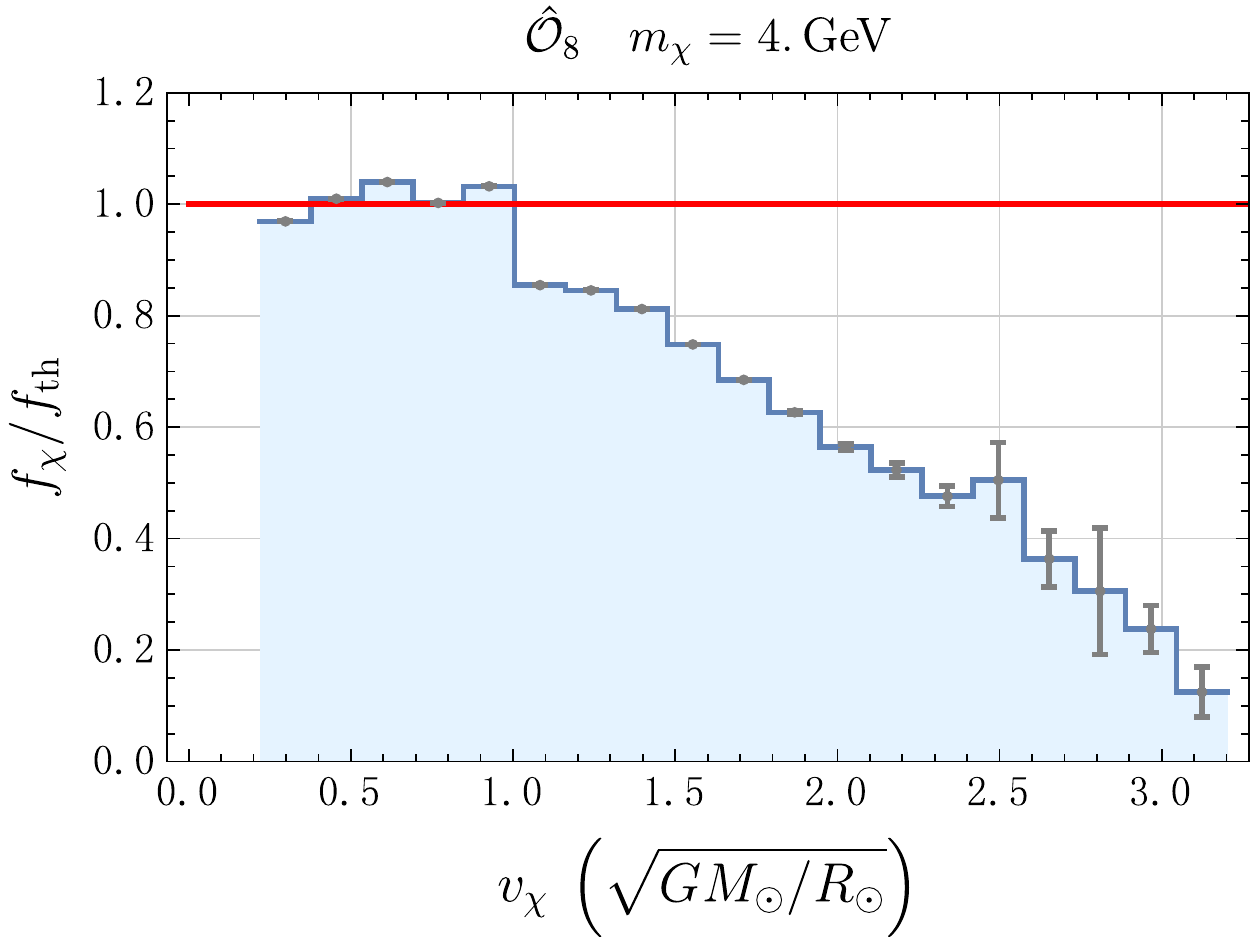}$\quad$\includegraphics[scale=0.57]{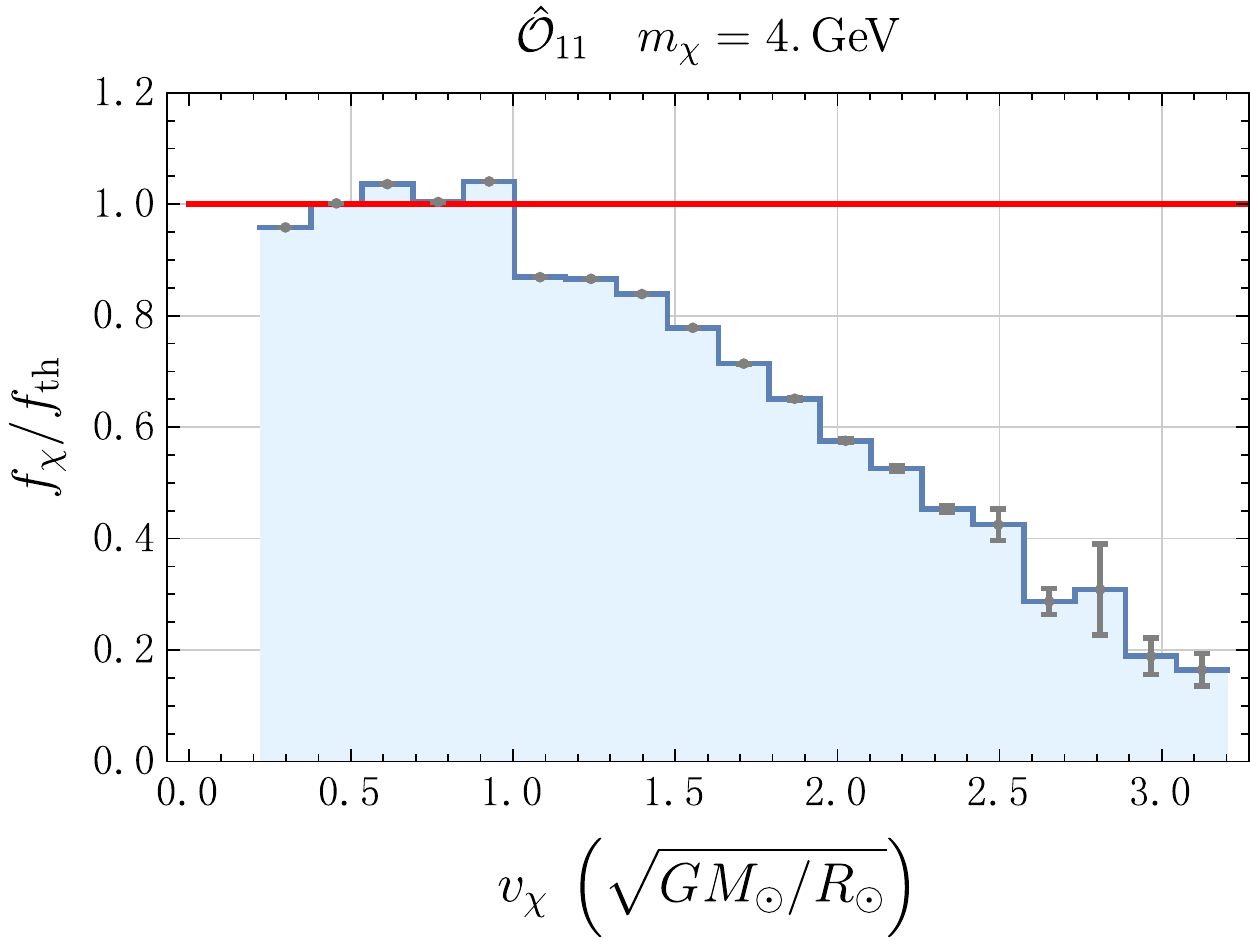}
\par\end{centering}

\begin{centering}
\vspace{0.1cm}

\par\end{centering}

\begin{centering}
\includegraphics[scale=0.57]{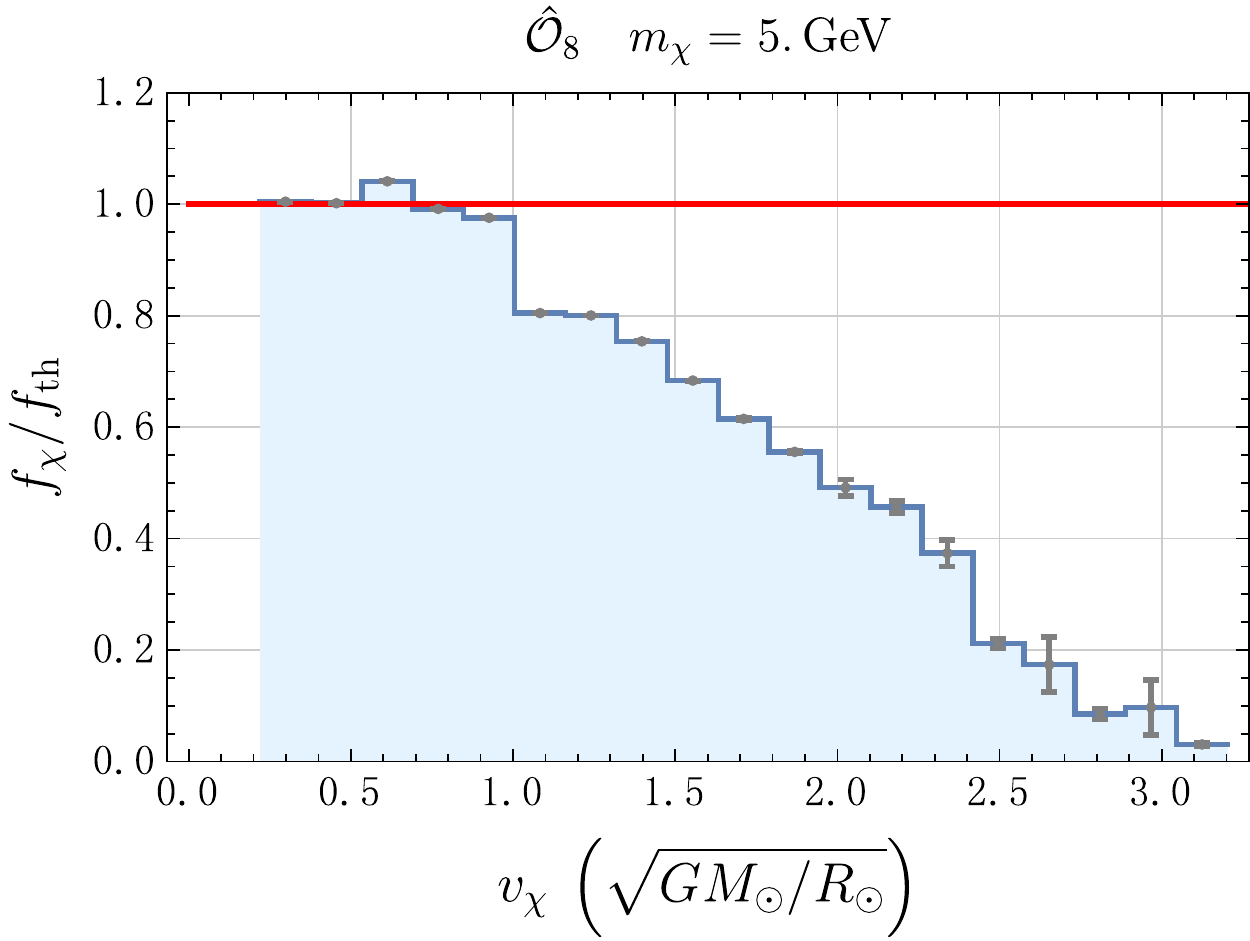}$\quad$\includegraphics[scale=0.57]{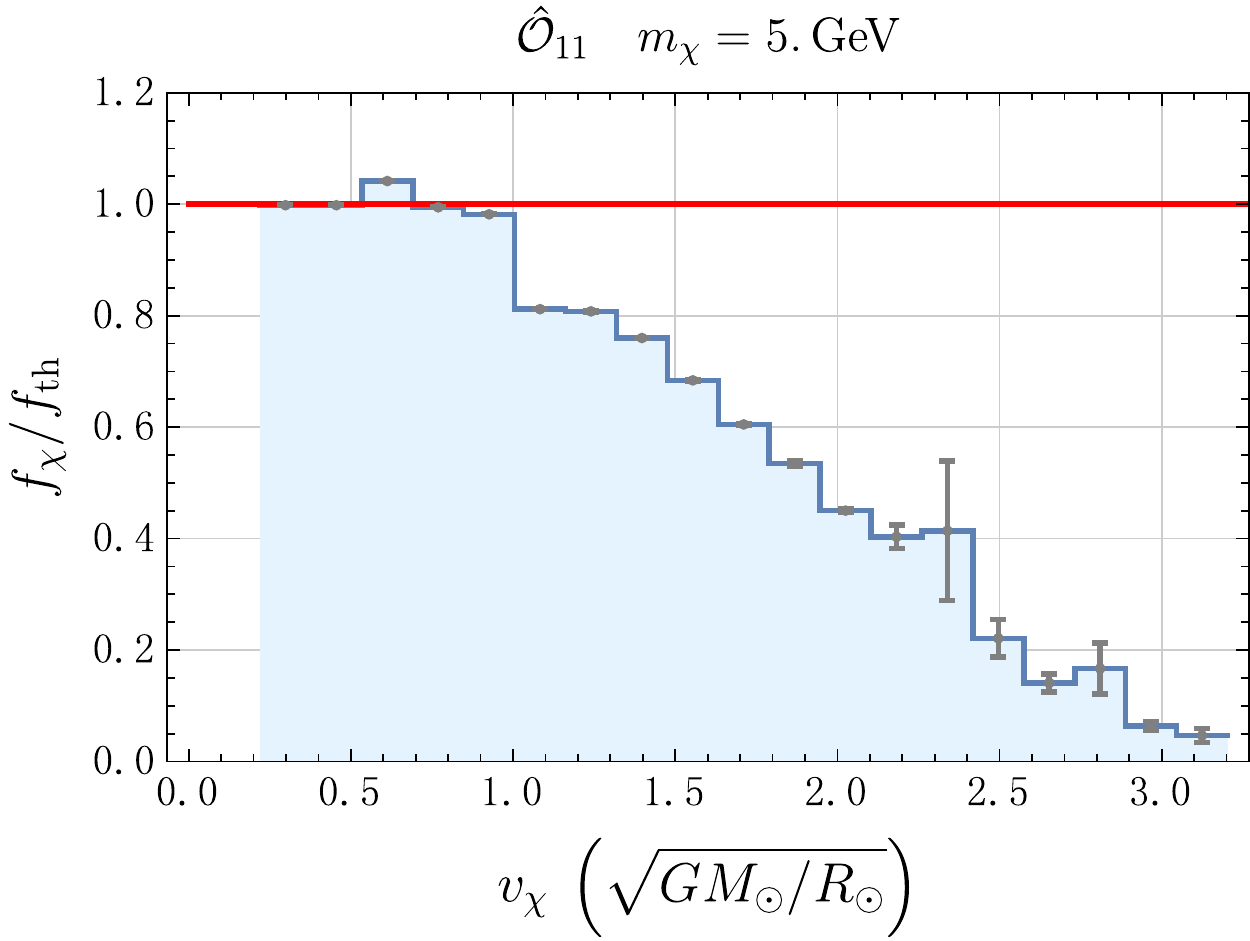}
\par\end{centering}

\protect\caption{\label{fig:RatioVelocity2}Parallel to Fig.~\ref{fig:RatioVelocity1}
for effective operators $\hat{\mathcal{O}}_{8}$ ($left$) and $\hat{\mathcal{O}}_{11}$
($right$), respectively. }
\end{figure}
Although the Maxwellian form of DM velocity distribution fails to
describe the tail of the actual velocity distribution, as mentioned
in Sec.~\ref{sec:Introduction}, it suffices to approximate the bulk
of the non-thermal distribution, on which physical processes such
as DM annihilation can be evaluated easily and accurately. The approximate
thermal distribution is expressed as $f_{\mathrm{th}}\propto\exp\left(-m_{\chi}E/T_{\chi}\right)$,
with the effective temperature parameter
$T_{\chi}$. $T_{\chi}$ is determined by the demand that there be
$no$ net energy transfer from the solar nuclei to the shuttling DM
particles once the steady state has been achieved, a requirement corresponds
to the following energy-moment equation~\cite{Spergel:1984re}:
\begin{eqnarray}
\int_{0}^{R_{\odot}}n_{\mathcal{A}}\left(r\right)\left[\frac{m_{\mathcal{A}}T_{\chi}+m_{\chi}T_{\odot}\left(r\right)}{m_{\mathcal{A}}\, m_{\chi}}\right]^{1/2}\left[T_{\odot}\left(r\right)-T_{\chi}\right]e^{-\frac{m_{\chi}V\left(r\right)}{T_{\chi}}}r^{2}dr & = & 0,
\end{eqnarray}
where $m_{\mathcal{A}}$ and $n_{\mathcal{A}}\left(r\right)$ are
the mass and the local number density of element $\mathcal{A}$, $T_{\odot}\left(r\right)$
is the temperature within the Sun, and $V\left(r\right)$ is the gravitational
potential as the function of radius $r$. In Tab.~\ref{tab:Xeff}
shown is the effective temperature $T_{\chi}$ for some benchmark
DM masses from $1$ $\mathrm{GeV}$ to $100$ $\mathrm{GeV}$. For
a DM particle weighing tens of GeV, the effective temperature $T_{\chi}$
can be approximated as the solar centre temperature $T_{\odot}\left(0\right)$.
\begin{table}
\begin{centering}
\begin{tabular}{clcc|cccc}
 &  &  & \multicolumn{1}{c}{} &  &  &  & \tabularnewline
\hline
\hline
$m_{\chi}\,\left(\mathrm{GeV}\right)$ & $\qquad$ & \multirow{1}{*}{$T_{\chi}/T_{\odot}\left(0\right)$} &  &  & $m_{\chi}\,\left(\mathrm{GeV}\right)$ & $\qquad$ & $T_{\chi}/T_{\odot}\left(0\right)$\tabularnewline
\hline
1 & $\qquad$ & 0.789 & $\;$ & $\;$ & 8 & $\qquad$ & 0.958\tabularnewline
2 & $\qquad$ & 0.867 &  &  & 9 & $\qquad$ & 0.962\tabularnewline
3 & $\qquad$ & 0.903 &  &  & 10 & $\qquad$ & 0.966\tabularnewline
4 & $\qquad$ & 0.923 &  &  & 15 & $\qquad$ & 0.977\tabularnewline
5 & $\qquad$ & 0.937 &  &  & 20 & $\qquad$ & 0.982\tabularnewline
6 & $\qquad$ & 0.946 &  &  & 50 & $\qquad$ & 0.993\tabularnewline
\multicolumn{1}{c}{7} & \multicolumn{1}{l}{$\qquad$} & 0.952 &  &  & 100 & $\qquad$ & 0.996\tabularnewline
\hline
\hline
 &  &  & \multicolumn{1}{c}{} &  &  &  & \tabularnewline
\end{tabular}
\par\end{centering}

\protect\caption{\label{tab:Xeff} Effective temperature $T_{\chi}$ for DM mass $m_{\chi}$
ranging from $1$ $\mathrm{GeV}$ to $100$ $\mathrm{GeV}$.}

\end{table}

For contrast, we compare the simulated velocity distribution $f_{\chi}$
to the approximate thermal one $f_{\mathrm{th}}$ in Fig.~\ref{fig:RatioVelocity1}
and Fig.~\ref{fig:RatioVelocity2} for effective operators $\hat{\mathcal{O}}_{1}$,
$\hat{\mathcal{O}}_{5}$, $\hat{\mathcal{O}}_{8}$ and $\hat{\mathcal{O}}_{11}$
in terms of the ratio $f_{\chi}/f_{\mathrm{th}}$. To estimate
the errors that propagate from  the simulated
scattering matrices, we  also present the standard deviations of the
discrete limiting distributions for each set of parameters~($\hat{\mathcal{O}}_{i}$,
$m_{\chi}$) in Fig.~\ref{fig:RatioVelocity1}
and Fig.~\ref{fig:RatioVelocity2}. Since in simulation  the transitions are restricted to only
the bound states, the DM velocity $v_{\chi}$ stretches to no further
than the escape velocity at the solar core $v_{\mathrm{esc}}\left(0\right)\approx3.17$. Echoing the
studies in Refs.~\cite{Nauenberg:1986em,Gould:1987ju},  while the ratio  $f_{\chi}/f_{\mathrm{th}}$ turns out to be  suppressed  at the high end of the velocity distribution, such suppression tends to be more significant for larger DM masses.

\subsection{\label{sub:evaporation}evaporation, capture and the minimum testable mass of the solar DM}

In Ref.~\cite{Gould:1987ju}, the author provided a thorough discussion
on the DM evaporation, under the assumption of a constant DM-nucleon
cross section, which corresponds to the operator $\hat{\mathcal{O}}_{1}$
in the context of the effective operators. Now we extends the discussion
to include other SI effective operators $\hat{\mathcal{O}}_{5}$,
$\hat{\mathcal{O}}_{8}$ and $\hat{\mathcal{O}}_{11}$. Our interest are focused on the scenario in which the Sun is optically thin to the DM particles, so an evaporation event is counted once the speed of scattered
DM particle exceeds the local escape velocity. For large DM-nucleus
cross section, the blocking effect due to multiple collisions has
to be taken into consideration, which turns out to heavily suppress
the evaporation~\cite{1990ApJ...356..302G}. However, as will be
shown later, the DM direct detections disfavour
the coupling parameters relevant for the optically thick regime for these SI effective operators. As a consequence,
a large optical depth for the solar DM particles  amounts to a satisfactory approximation within the scope of this
work.

Following Ref.~\cite{Gould:1987ju}, we start with the quantity $R_{\mathcal{A}}\left(w\rightarrow v\right)$
which represents the possibility of a DM particle with initial velocity
$w$ scattered to final velocity $v$ by nucleus $\mathcal{A}$ in
a unit volume,
\begin{eqnarray}
R_{\mathcal{A}}\left(w\rightarrow v\right) & = & n_{\mathcal{A}}\left\langle \frac{\mathrm{d}\sigma_{\chi\mathcal{A}}\left(\left|\mathbf{w}-\mathbf{u}_{\mathcal{A}}\right|\right)}{\mathrm{d}v}\left|\mathbf{w}-\mathbf{u}_{\mathcal{A}}\right|\right\rangle ,\nonumber \\
 & = & n_{\mathcal{A}}\int f_{\mathcal{A}}\left(\mathbf{u}_{\mathcal{A}}\right)\frac{\mathrm{d}\sigma_{\chi\mathcal{A}}\left(\left|\mathbf{w}-\mathbf{u}_{\mathcal{A}}\right|\right)}{\mathrm{d}v}\,\left|\mathbf{w}-\mathbf{u}_{\mathcal{A}}\right|\mathrm{d}^{3}u_{\mathcal{A}}\label{eq:R1}
\end{eqnarray}
where $\mathrm{d}\sigma_{\chi\mathcal{A}}\left(\left|\mathbf{w}-\mathbf{u}_{\mathcal{A}}\right|\right)/\mathrm{d}v$
is the differential cross section for the DM-nucleus system, which
depends on their relative velocity $\mathbf{w}-\mathbf{u}_{\mathcal{A}}$,
and $\left\langle \cdots\right\rangle $ denotes the average over
the thermal velocity distribution of element $\mathcal{A}$. The Maxwellian
distribution $f_{\mathcal{A}}\left(\mathbf{u}_{\mathcal{A}}\right)$
is written as
\begin{eqnarray}
f_{\mathcal{A}}\left(\mathbf{u}_{\mathcal{A}}\right) & = & \left(\sqrt{\pi}u_{0}\right)^{-3}\exp\left(-\frac{u_{\mathcal{A}}^{2}}{u_{0}^{2}}\right),\label{eq:maxwellian}
\end{eqnarray}
where $u_{0}=$ $\sqrt{2\, T_{\odot}/m_{\mathcal{A}}}$. For the purpose
of concision, we postpone the explicit expression of Eq.~(\ref{eq:R1})
to Appendix~\ref{sec:appendixb}. Next, given DM velocity $w$ and
the escape velocity $v_{\mathrm{esc}}$, the evaporation rate in the
unit volume can be written as
\begin{eqnarray}
\varOmega^{+}\left(w\left|v_{\mathrm{esc}}\right.\right) & = & \sum_{\mathcal{A}}\int_{v_{\mathrm{esc}}}^{+\infty}R_{\mathcal{A}}\left(w\rightarrow v\right)\mathrm{d}v,
\end{eqnarray}
where the summation is taken over all solar elements. Finally, by
convoluting $\varOmega^{+}\left(w\left|v_{\mathrm{esc}}\right.\right)$
with DM distribution $f_{\chi}\left(r,\, w\right)$ determined from
simulation, we express the DM evaporation rate as follows
\begin{eqnarray}
E_{\odot} & = & \int\varOmega^{+}\left(w\left|v_{\mathrm{esc}}\right.\right)f_{\chi}\left(r,\, w\right)\mathrm{d}r\,\mathrm{d}w,
\end{eqnarray}
where $\varOmega^{+}\left(w\left|v_{\mathrm{esc}}\right.\right)$
depends on the radial coordinate $r$ through the distributions of
solar nuclei and the escape velocity $v_{\mathrm{esc}}\left(r\right)$,
which are both described with the SSM GS98 \cite{Serenelli:2009yc}.
Given $j_{\chi}=1/2$, the evaporation rate for various SI effective
operators are expressed with the following fitting functions:
\begin{subequations}
\begin{eqnarray}
E_{\odot}^{\hat{\mathcal{O}}_{1}} & \simeq & 1.49\times10^{-2.63\,\left[\left(\frac{m_{\chi}}{1\,\mathrm{GeV}}\right)^{1.11}+\left(\frac{1\,\mathrm{GeV}}{m_{\chi}}\right)^{-0.03}\right]}\left(\frac{\sigma_{\mathrm{p}}}{10^{-40}\,\mathrm{cm}^{2}}\right)10^{-4}\,\mathrm{s}^{-1},\label{eq:EvaporationO1}\\
E_{\odot}^{\hat{\mathcal{O}}_{5}} & \simeq & 2.01\times10^{-1.92\,\left[\left(\frac{m_{\chi}}{1\,\mathrm{GeV}}\right)^{1.23}+\left(\frac{1\,\mathrm{GeV}}{m_{\chi}}\right)^{-0.07}\right]}\left(\frac{c_{5}}{10^{-1}\,\mathrm{GeV}^{-2}}\right)^{2}10^{-6}\,\mathrm{s}^{-1}\\
E_{\odot}^{\hat{\mathcal{O}}_{8}} & \simeq & 4.08\times10^{-2.41\,\left[\left(\frac{m_{\chi}}{1\,\mathrm{GeV}}\right)^{1.17}+\left(\frac{1\,\mathrm{GeV}}{m_{\chi}}\right)^{-0.25}\right]}\left(\frac{c_{8}}{10^{-3}\,\mathrm{GeV}^{-2}}\right)^{2}10^{-5}\,\mathrm{s}^{-1}\\
E_{\odot}^{\hat{\mathcal{O}}_{11}} & \simeq & 1.82\times10^{-1.77\,\left[\left(\frac{m_{\chi}}{1\,\mathrm{GeV}}\right)^{1.26}+\left(\frac{1\,\mathrm{GeV}}{m_{\chi}}\right)^{-0.02}\right]}\left(\frac{c_{11}}{10^{-4}\,\mathrm{GeV}^{-2}}\right)^{2}10^{-7}\,\mathrm{s}^{-1},
\end{eqnarray}
\end{subequations}
which approximate the numerical results with an accuracy better than
$10\%$ in the DM mass range  $2\leq m_{\chi}\leq5\,\mathrm{GeV}$.
For the sake of convenience, we invoke the DM-nucleon cross section
$\sigma_{\mathrm{p}}=c_{1}^{2}\,\mu_{N}^{2}/\pi$ instead of coupling
parameter $c_{1}$ in Eq.~(\ref{eq:EvaporationO1}).

Here we take a short review of the solar capture rate $C_{\odot}$
and the annihilation coefficient $A_{\odot}$. The standard procedure
for evaluating the DM capture rate $C_{\odot}$ is developed in the
literature~\cite{Gould:1987ir,Gould:1987ww,Gould:1991hx}. Given
the Galactic DM distribution unperturbed by solar influence, we first
derive the collision event rate using the Liouville theorem and angular
momentum conservation in the solar central force field, and  by
demanding the momentum transfer be large enough for the capture, we
then extract the capture rate out of the total collision event rate. While
discussions on capture rates for various DM-nucleon effective operators
can be found in Refs.~\cite{Liang:2013dsa,Catena:2015uha}, here
we present the numerical results for $j_{\chi}=1/2$ in the DM mass
range $2\,\mathrm{GeV}\leq m_{\chi}\leq5\,\mathrm{GeV}$ as the following
fitting functions dependent on the DM mass $x=\left(m_{\chi}/1\,\mathrm{GeV}\right)$:
\begin{subequations}
\begin{eqnarray}
C_{\odot}^{\hat{\mathcal{O}}_{1}} & \simeq & \left(-1.17023+17.9214\, x-15.0294\, x^{2}+6.30696\, x^{3}-1.43792\, x^{4}+0.170425\, x^{5}\right.\nonumber \\
 &  & \left.-0.008241\, x^{6}\right)\left(\frac{\sigma_{\mathrm{p}}}{10^{-40}\,\mathrm{cm}^{2}}\right)10^{25}\,\mathrm{s}^{-1},
\end{eqnarray}
\begin{eqnarray}
C_{\odot}^{\hat{\mathcal{O}}_{5}} & \simeq & \left(6.73314-12.5207\, x+9.48633\, x^{2}-3.63890\, x^{3}+0.771875\, x^{4}-0.0849675\, x^{5}\right.\nonumber \\
 &  & \left.+0.00379191\, x^{6}\right)\left(\frac{c_{5}}{10^{-1}\,\mathrm{GeV}^{-2}}\right)^{2}10^{26}\,\mathrm{s}^{-1}
\end{eqnarray}
\begin{eqnarray}
C_{\odot}^{\hat{\mathcal{O}}_{8}} & \simeq & \left(6.33402-11\text{.}3047\, x+8.86278\, x^{2}-3.54797\, x^{3}+0.775692\, x^{4}-0.0882098\, x^{5}\right.\nonumber \\
 &  & \left.+0.00408605\, x^{6}\right)\left(\frac{c_{8}}{10^{-3}\,\mathrm{GeV}^{-2}}\right)^{2}10^{26}\,\mathrm{s}^{-1}
\end{eqnarray}
\begin{eqnarray}
C_{\odot}^{\hat{\mathcal{O}}_{11}} & \simeq & \left(4.69007-8.90451\, x+6.98704\, x^{2}-2.69955\, x^{3}+0.592700\, x^{4}-0.0683059\, x^{5}\right.\nonumber \\
 &  & \left.+0.00322178\, x^{6}\right)\left(\frac{c_{11}}{10^{-4}\,\mathrm{GeV}^{-2}}\right)^{2}10^{25}\,\mathrm{s}^{-1}.
\end{eqnarray}
\end{subequations}In above evaluation of the  capture rates, we adopt the isothermal DM
halo model with a local density $\rho_{\chi}=0.3\,\mathrm{GeV\cdot cm^{-3}}$
and a Maxwellilan velocity distribution with the dispersion $v_{0}=220\,\mathrm{km\cdot s^{-1}}$,
truncated at the Galactic escape velocity of $544\,\mathrm{km\cdot s^{-1}}$.

The annihilation coefficient $A_{\odot}$ can be expressed in terms
of the thermal cross section $\left\langle \sigma v\right\rangle _{\odot}$
and the effective occupied volume of the solar DM $V_{\mathrm{eff}}$
as the following:
\begin{eqnarray}
A_{\odot} & \equiv & \frac{\left\langle \sigma v\right\rangle _{\odot}}{V_{\mathrm{eff}}},
\end{eqnarray}
and the effective volume can be described with the fitting function

\begin{equation}
V_{\mathrm{eff}}=6.9\times10^{27}\left(\frac{100\,\mathrm{GeV}}{m_{\chi}}\right)^{3/2}\,\mathrm{cm^{3}}.
\end{equation}

\begin{figure}
\begin{centering}
\includegraphics[scale=0.6]{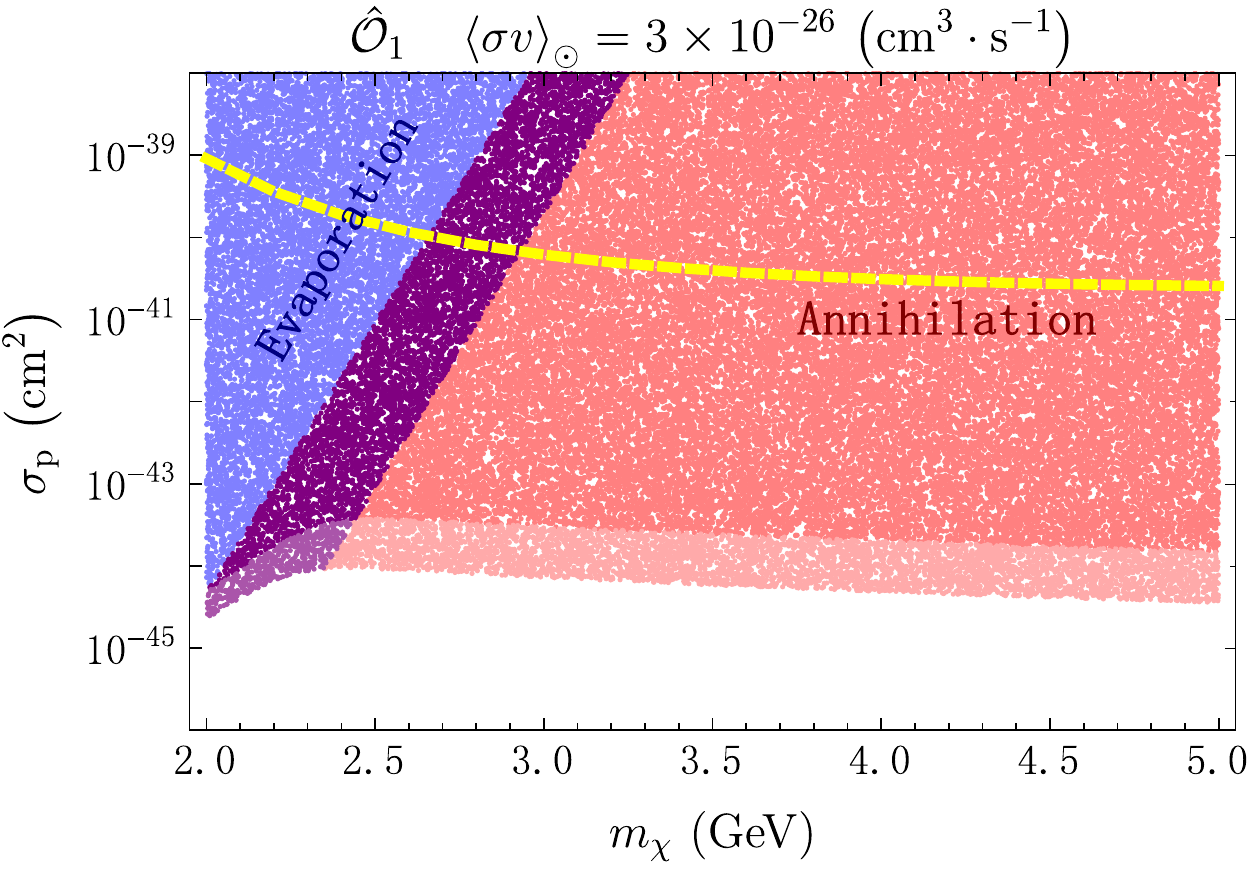}$\quad$\includegraphics[scale=0.6]{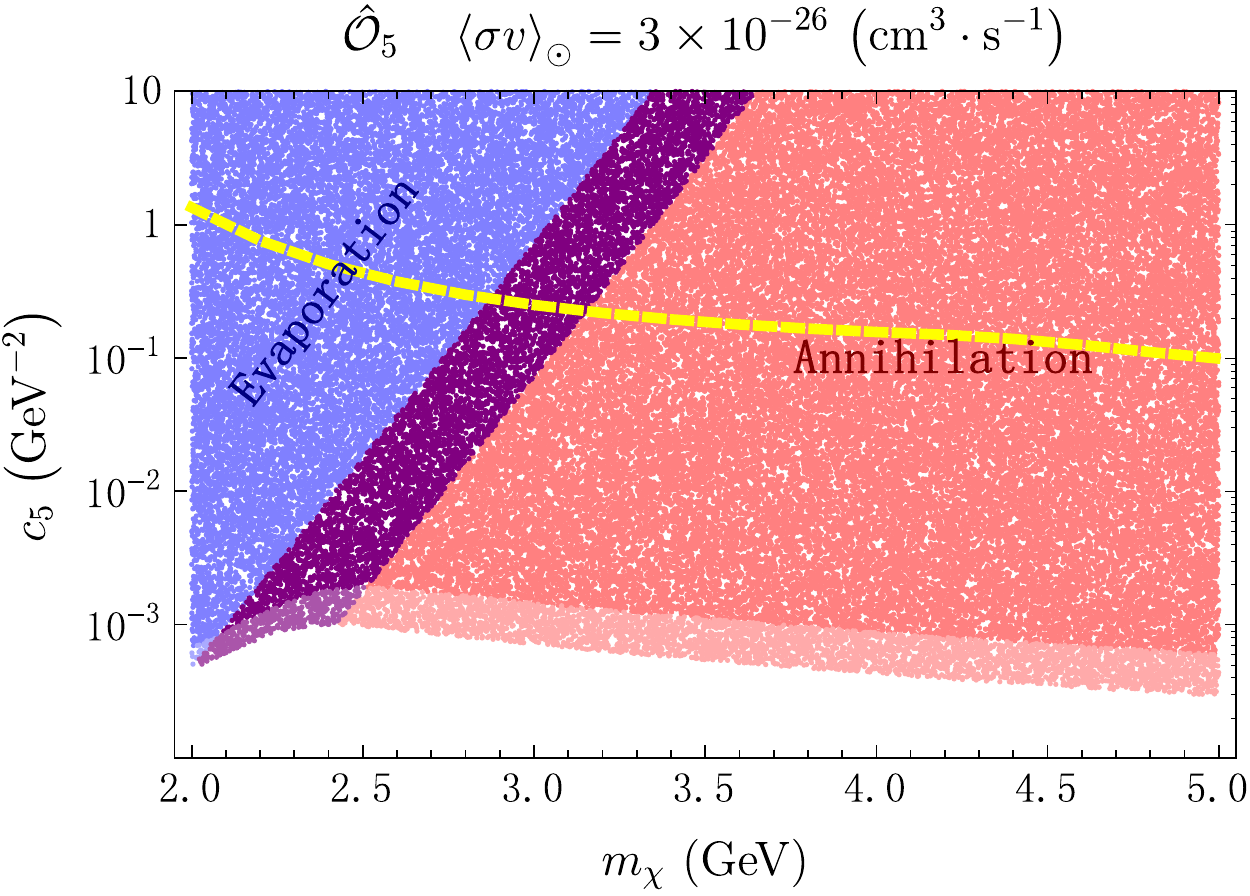}
\par\end{centering}

\begin{centering}
\includegraphics[scale=0.6]{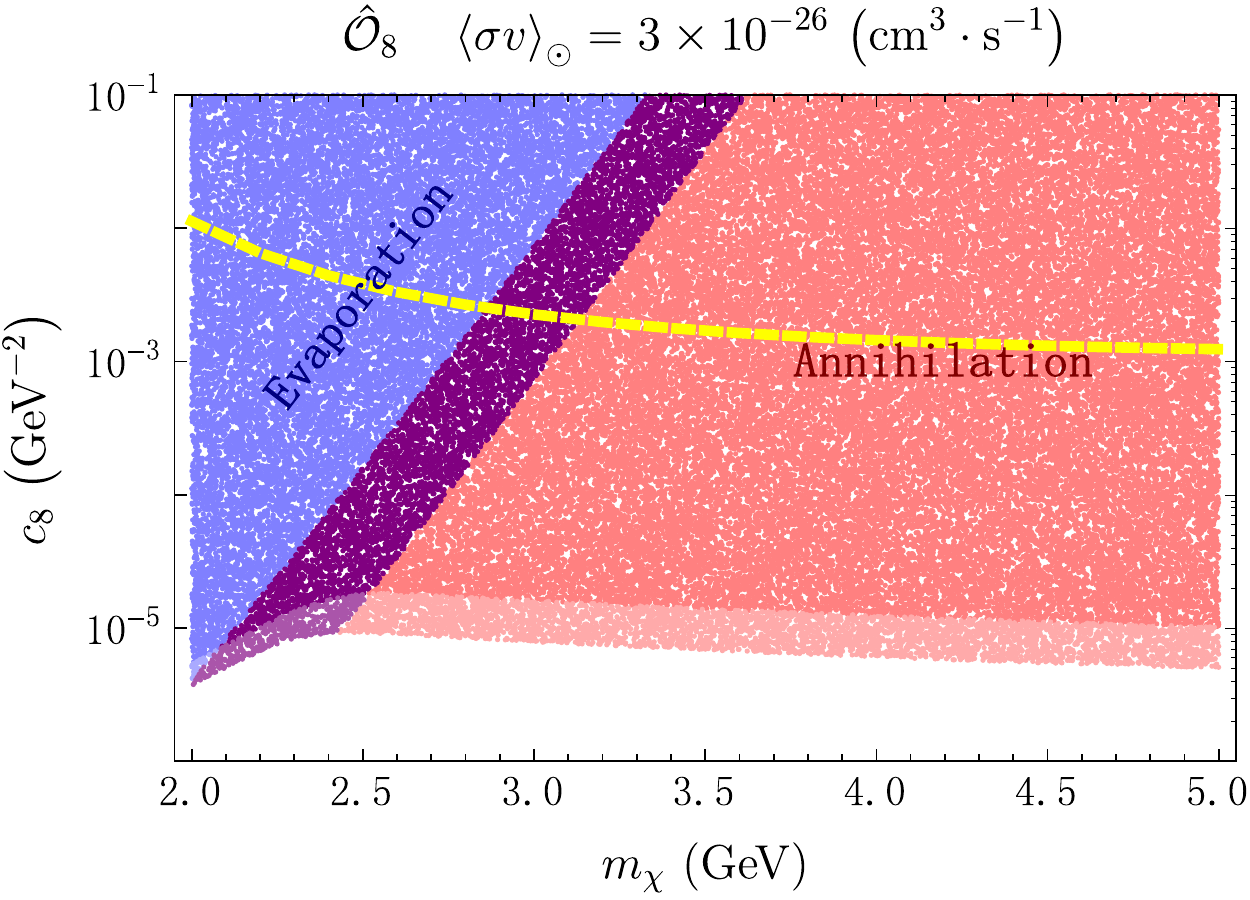}$\quad$\includegraphics[scale=0.6]{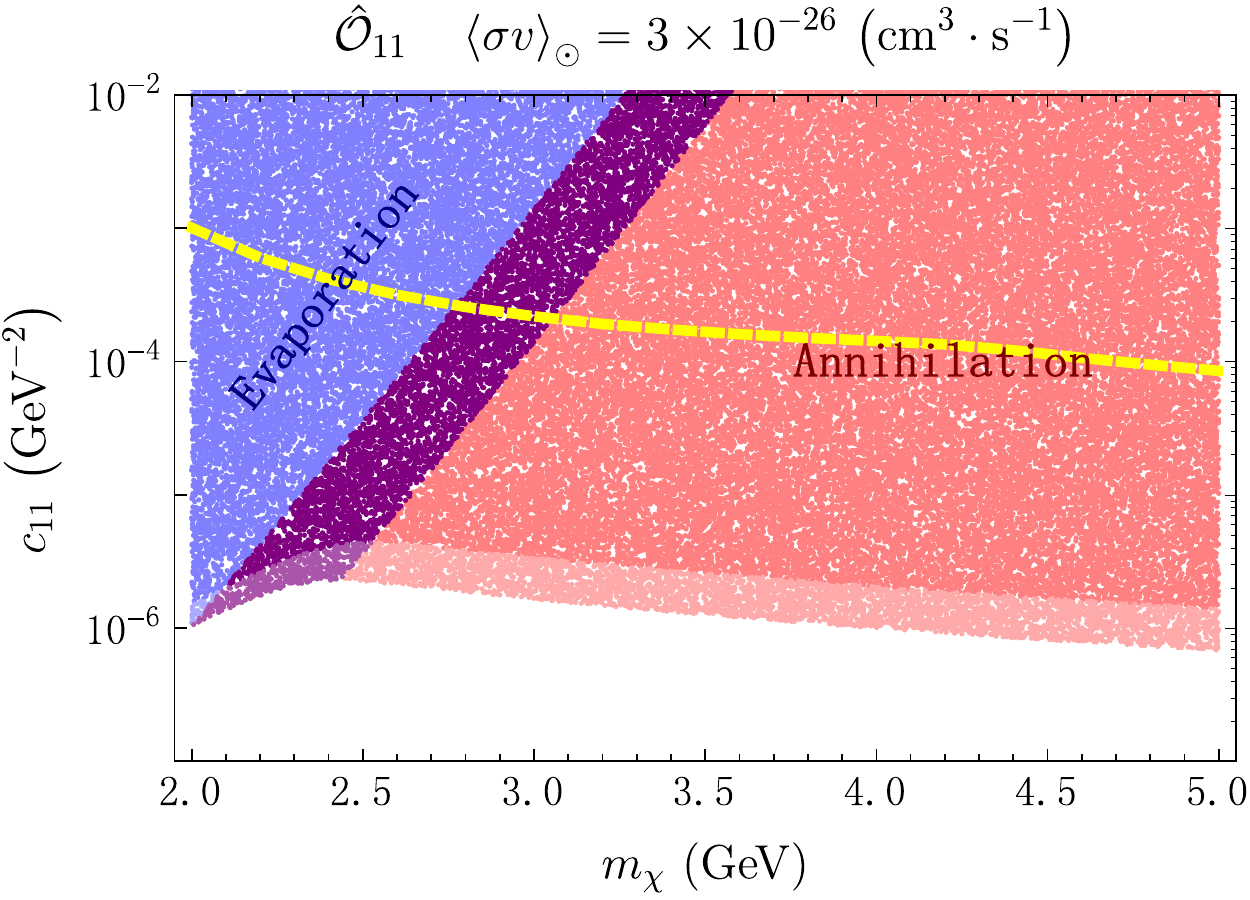}
\par\end{centering}

\protect\caption{\label{fig:parameterRegion}The parameter regions of DM mass and DM-nucleon
coupling strength for operators $\hat{\mathcal{O}}_{1}$, $\hat{\mathcal{O}}_{5}$,
$\hat{\mathcal{O}}_{8}$ and $\hat{\mathcal{O}}_{11}$ for $j_{\chi}=1/2$.
While the signal regions $\left(\tanh\left(t_{\odot}/\tau_{\mathrm{e}}\right)\simeq1\right)$
are presented as the darker-coloured areas, the lighter counterparts
correspond to the region where $0.9\leq\tanh\left(t_{\odot}/\tau_{\mathrm{e}}\right)\apprle1$ for reference.
In the red (blue) area, evaporation (annihilation) plays a sub-dominant
role in the evolution of the solar DM number. The purple belt represents
the transition zone where both evaporation and annihilation effects
are of equal importance. The 90\% C.L. upper
bounds (yellow dashed lines) are inferred from the binned data of the
CDMSlite~\cite{Agnese:2015nto}. See text for more details.}
\end{figure}

Now we are ready to explore the parameter space where the solar neutrino
observational approach is effective for the DM detection, putting
our intuitive discussion in Sec.~\ref{sec:Introduction} onto concrete
computation. On one hand, as mentioned in Sec.~\ref{sec:Introduction},
to ensure the full strength of the neutrino flux it is required that
$\tanh\left(t_{\odot}/\tau_{\mathrm{e}}\right)\simeq1$, for which
we adopt the criterion $t_{\odot}/\tau_{\mathrm{e}}\gtrsim3.0$. On
the other hand, to specify the parameter region for the annihilation-
and evaporation-dominated scenarios, we set the criteria as $E_{\odot}^{2}/\left(4\, C_{\odot}A_{\odot}\right)\leq0.1$
and $E_{\odot}^{2}/\left(4\, C_{\odot}A_{\odot}\right)\geq10$, respectively.
For concreteness, in Fig.~\ref{fig:parameterRegion} we show the
relevant parameter regions for the annihilation- and evaporation-dominated
regimes for SI effective operators $\hat{\mathcal{O}}_{1}$, $\hat{\mathcal{O}}_{5}$,
$\hat{\mathcal{O}}_{8}$ and $\hat{\mathcal{O}}_{11}$, by assuming
the canonical $s$-wave thermal annihilation cross section $\left\langle \sigma v\right\rangle _{\odot}=3\times10^{-26}\,\mathrm{cm}^{2}$,
although the $p$-wave annihilation is also possible. Also shown in Fig.~\ref{fig:parameterRegion}
(in yellow dashed lines) are the 90\% C. L. upper limits on the DM-nucleon
couping strengths imposed by the second run of the CDMSlite~\cite{Agnese:2015nto},
which are derived using the Poisson statistics based on the event
spectrum, signal efficiency, and detector resolution presented in
Ref.~\cite{Agnese:2015nto}, along with the astrophysical parameters
consistent with the calculation of the capture rate. The new CDMSlite
constraints are strong enough for narrowing our investigation to the
optically thin regime. To illustrate this, taking  $\hat{\mathcal{O}}_{1}$  for example, we note that the upper bound of  $\sigma_{\mathrm{p}}\approx10^{-39}\,\mathrm{cm}^{2}$
corresponds to a mean free path at the solar centre  $l_{\chi}\left(0\right)=\left(\ensuremath{\sum\limits _{\mathcal{A}}}n_{\mathcal{A}}\left(0\right)\sigma_{\chi\mathcal{A}}\right)^{-1}\approx10\, R_{\odot}$, with  $\sigma_{\chi\mathcal{A}}$   the DM-nucleus cross section. So
 the assumption of a large optical depth is justified.

Given above quantitative analysis, we are able to draw clear boundaries
among different signal topologies. For instance, for the effective
interaction $\hat{\mathcal{O}}_{1}$ with a DM-nucleon cross section
$\sigma_{\mathrm{p}}=10^{-40}\,\mathrm{cm}^{2}$, the assumption of
an equilibrium between capture and annihilation is only valid for
a DM particle heavier than $2.96\,\mathrm{GeV}$, while for a DM mass
smaller than $2.67\,\mathrm{GeV}$, one can no longer  extract
the coupling strength of the DM-nucleon interaction from the observed
neutrino flux, because the number of DM particles $N_{\chi}=C_{\odot}/E_{\odot}$
becomes independent of cross section $\sigma_{\mathrm{p}}$~\cite{Busoni:2013kaa}\footnote[4]{\renewcommand{\baselinestretch}{3}\selectfont The detection of the solar DM evaporation is discussed in  Ref.~\cite{Kouvaris:2015nsa}.}.
In addition, if the DM-nucleon cross section $\sigma_{\mathrm{p}}$ is smaller roughly
than $10^{-44}\,\mathrm{cm}^{2}$, the equilibrium among capture,
evaporation and annihilation has not yet been achieved at the present
day. As a consequence, the signal flux is suppressed and the unsaturated
number of the solar DM (Eq.~(\ref{eq:DMnumber})) needs to be specified
for  neutrino telescopes to determine or constrain the coupling strength
(see, $e.g.$, Ref.~\cite{Aartsen:2016exj}).

\section{\label{sec:Conclusion}Discussions }

As mentioned in Sec.~\ref{sec:Introduction}, authors of Refs.~\cite{Vincent:2014jia,Vincent:2015gqa}
introduce the weakly interacting asymmetric dark matter (ADM) with
generalised form factors in an attempt to solve the solar abundance
problem. Without annihilation the ADM may accumulate to such amount
that their presence can slightly affect the solar structure. Assuming
the evaporation rate is zero, it is found that the following SI interaction
between a $3\,\mathrm{GeV}$ ADM and nucleon gives the best result:
\begin{eqnarray}
\sigma & = & \sigma_{0}\left(\frac{q}{q_{0}}\right)^{2},\label{eq:FitModel}
\end{eqnarray}
where the coupling $\sigma_{0}=10^{-37}\,\mathrm{cm}^{2}$, and the
reference momentum $q_{0}=40\,\mathrm{MeV}$. The translation between
the contexts of the generalised form factor and the effective operator $\hat{\mathcal{O}}_{11}$
is realised through the relation
\begin{eqnarray}
\frac{\pi}{\mu_{N}^{2}}\sigma_{0}\left(\frac{q}{q_{0}}\right)^{2} & = & \frac{j_{\chi}\left(j_{\chi}+1\right)}{3\,}c_{11}^{2}\left(\frac{q}{m_{N}}\right)^{2},
\end{eqnarray}
which gives $c_{11}=1.87\times10^{-3}\,\mathrm{GeV}^{-2}$ for $j_{\chi}=1/2$.

\begin{figure}
\begin{centering}
\includegraphics[scale=0.6]{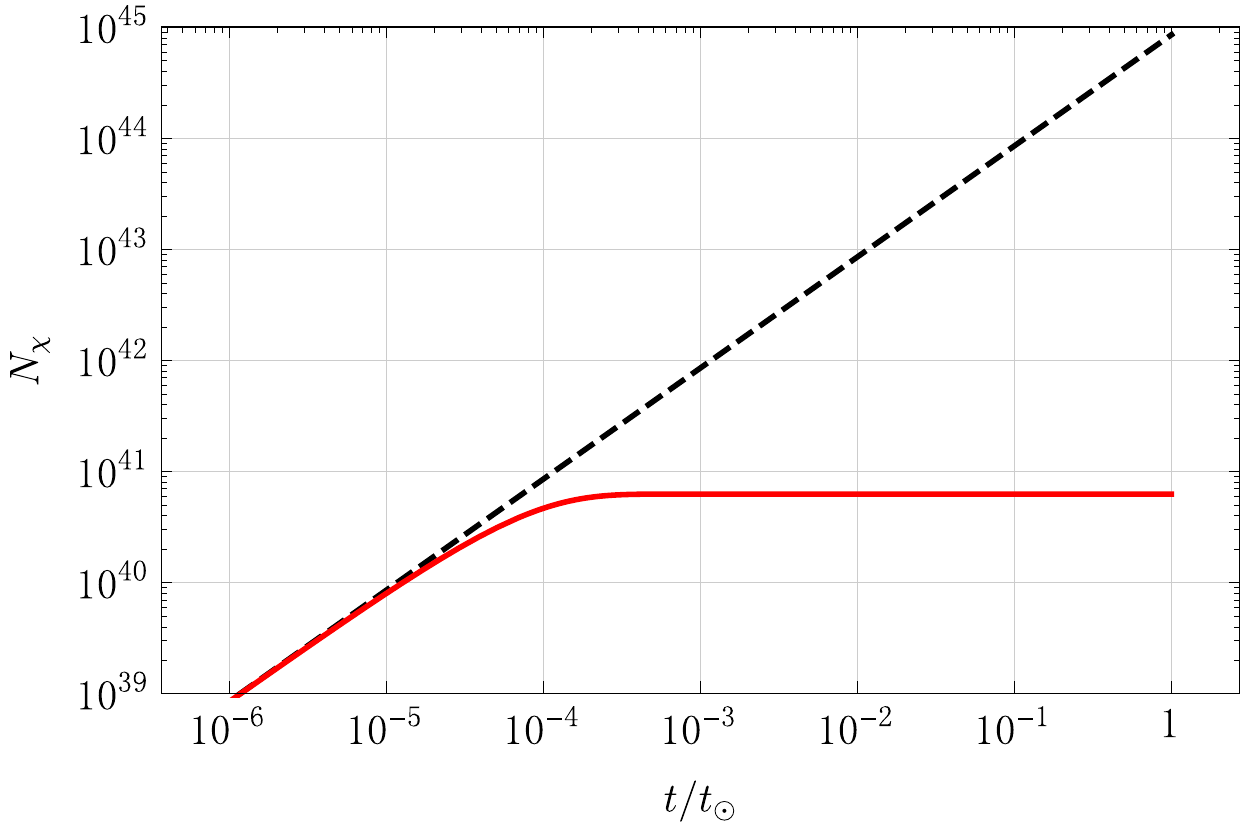}\protect\caption{\label{fig:ADMNumber}The number of the solar DM with $\left(red\right)$
and without $\left(black\, dashed\right)$ evaporation for parameters
$m_{\chi}=3\,\mathrm{GeV}$, $\sigma_{0}=10^{-37}\,\mathrm{cm}^{2}$,
and $q_{0}=40\,\mathrm{MeV}$.}

\par\end{centering}

\end{figure}

For the best-fit parameters given above, we calculate the evolution of
the solar DM $with$ and $without$ evaporation in Fig.~\ref{fig:ADMNumber}.
It is evident that the presence of evaporation significantly constrain
the increment of the DM number $N_{\chi}$ and freezes it at a number
$\mathcal{O}\left(10^{4}\right)$ smaller than the value without evaporation,
which indicates an inconsistency for the model in Eq.~(\ref{eq:FitModel})
to alleviate the discrepancies between the SSM and helioseismological
observables. Note that although we evaluate the evaporation rate by
neglecting the interplay between the accumulated DM population and
solar nuclei background, our calculation still holds in the ADM scenario
because the relevant effects only result in minor changes in the solar
structure. It should be also note that such inconsistency has been
confirmed by the DM direct detection  from the experimental aspect: $\mathrm{CRESST}$-\uppercase\expandafter{\romannumeral2}
 ruled out this particular model at  $90\%$ C.L.~\cite{Angloher:2016jsl}. In order to evade the constraints from the direct detection, the same authors of Refs.~\cite{Vincent:2014jia,Vincent:2015gqa} recently propose a spin-dependent (SD) $v^{2}$ interaction as an  alternative solution in Ref.~\cite{Vincent:2016dcp}. We leave the discussion on the relevant evaporation effect in the SD scenario for future work.

Finally, we discuss a subtlety underlying the methodology applied
to calculate the steady distribution $f_{\chi}\left(E,\, L\right)$
in Sec.~\ref{sec:Distribution=000026Evaporation}, $i.e.$, to what extent
the Markov chain approach  describes the realistic evolution
of the solar DM distribution, considering that both the replenishment
and the leakage of DM particles are not reflected in the master equation
Eq.~(\ref{eq:master equation}). To this end, we explicitly write
down the differential increment of the solar DM number in a time step
$\delta t$,
\begin{eqnarray}
N_{\chi}\left(t+\delta t\right)\boldsymbol{\xi}' & = & C_{\odot}\,\delta t\,\boldsymbol{\eta}+N_{\chi}\left(t\right)\mathcal{S}\cdot\boldsymbol{\xi}-N_{\chi}\left(t\right)\delta t\,\mathcal{E\cdot\boldsymbol{\xi}},\label{eq:masterEquation}
\end{eqnarray}
where vector $\boldsymbol{\xi}^{T}=\left(\xi_{1},\,\xi_{2},\cdots,\,\xi_{n}\right)$
and $\boldsymbol{\xi}'^{T}=\left(\xi'_{1},\,\xi'_{2},\cdots,\,\xi'_{n}\right)$
denotes the normalised probability for the $n$ states at time $t$
and $t+\delta t$, respectively, and $\boldsymbol{\eta}^{T}=\left(\eta_{1},\,\eta_{2},\cdots,\,\eta_{n}\right)$
represents the distribution for the newly captured DM particles in
time interval $\delta t$. The Markov transition matrix $\mathcal{S}$
is expressed as
\begin{eqnarray}
\mathcal{S} & = & \left(\begin{array}{cccc}
1-\ensuremath{\sum\limits _{i\neq1}}S_{i1} & S_{12} & \cdots & S_{1n}\\
S_{21} & 1-\ensuremath{\sum\limits _{i\neq2}}S_{i2} & \cdots & \vdots\\
\vdots & \vdots & \ddots & \vdots\\
S_{n1} & S_{n2} & \cdots & 1-\ensuremath{\sum\limits _{i\neq n}}S_{in}
\end{array}\right),
\end{eqnarray}
with element $S_{ji}$ being the probability for the transition $i\rightarrow j$.
Matrix

\begin{eqnarray}
\mathcal{E} & = & \left(\begin{array}{cccc}
e_{1} & 0 & \cdots & 0\\
0 & e_{2} & \cdots & \vdots\\
\vdots & \vdots & \ddots & \vdots\\
0 & 0 & \cdots & e_{n}
\end{array}\right)
\end{eqnarray}
describes the leakage due to evaporation, with $e_{i}$ being the
evaporation rate for the $i$-th state. It is evident from Eq.~(\ref{eq:masterEquation})
that the equilibrium distribution of the Markov chain $\boldsymbol{\xi}_{\mathrm{eq}}$
which satisfies the equation $\mathcal{S}\cdot\boldsymbol{\xi}_{\mathrm{eq}}=\boldsymbol{\xi}_{\mathrm{eq}}$
well approximates the realistic distribution so long as the fractional
change of DM number is negligible in the relaxation time $\delta t=t_{\mathrm{relax}}$,
$i.e.$,
\begin{eqnarray}
\left|\frac{N_{\chi}\left(t+t_{\mathrm{relax}}\right)-N_{\chi}\left(t\right)}{N_{\chi}\left(t\right)}\right| & \ll & 1.
\end{eqnarray}
Therefore, for a time step $\delta t\gtrsim t_{\mathrm{relax}}$,
it is reasonable to assume that solar DM equilibrates to its limit
distribution instantaneously, and the descriptions of the distribution
and the total number of the solar DM decouple and thus can be treated
separately. Under such circumstance, one determines the evaporation
rate using the steady distribution function and in turn integrates
Eq.~(\ref{eq:master equation-1}) to obtain the number of the solar
DM in a self-consistent way. Note that for simplicity the annihilation
is not included in our discussion, which however, will not cause any
loss of generality of our conclusion.

\appendix

\begin{acknowledgments}
We thank Huang Da for helpful discussion on the CDMSlite constraints. This work is supported in part by the National Basic Research Program
of China (973 Program) under Grants No.~2010CB833000; the National
Nature Science Foundation of China (NSFC) under Grants No.~10905084,
No. 11335012 and No.~11475237; The numerical calculations were done
using the HPC Cluster of SKLTP/ITP-CAS.
\end{acknowledgments}

\begin{appendices}
\renewcommand\thesection{\Alph{section}}
\renewcommand{\theequation}{\Alph{section}.\arabic{equation}}
\section{\label{sec:Appendixa}collision probability }

As mentioned in Sec.~\ref{sub:sec.3.a}, we need to calculate the
collision probabilities in the time interval $\Delta t$ prior to
sampling the scattering events, and then as the weight these probabilities
are folded with the scattering samples so as to determine the transition
matrix $S\left(E,\, L;\, E',\, L'\right)$ in an efficient way. Here
we provide a brief discussion on the the collision probability.

Considering that the DM collision is described with the Poisson process,
the collision probability in time interval $\Delta t$ can be expressed
as
\begin{eqnarray}
P_{\mathrm{c}} & = & 1-\exp\left[-\int_{0}^{\Delta t}\lambda(\tau)\,\mathrm{d}\tau\right],\label{eq:collison P}
\end{eqnarray}
where\footnote[5]{\renewcommand{\baselinestretch}{1}\selectfont For simplicity we omit the summation notation over various solar elements $\mathcal{A}$.}
\begin{eqnarray}
\lambda & = & n_{\mathcal{A}}\left\langle \sigma\left(\left|\mathbf{w}-\mathbf{u}_{\mathcal{A}}\right|\right)\left|\mathbf{w}-\mathbf{u}_{\mathcal{A}}\right|\right\rangle \nonumber \\
 & = & n_{\mathcal{A}}\int f_{\mathcal{A}}\left(\mathbf{u}_{\mathcal{A}}\right)\sigma\left(\left|\mathbf{w}-\mathbf{u}_{\mathcal{A}}\right|\right)\,\left|\mathbf{w}-\mathbf{u}_{\mathcal{A}}\right|\mathrm{d}^{3}u_{\mathcal{A}}\label{eq:lambda}
\end{eqnarray}
is $implicitly$ dependent on time once the DM trajectory is determined.

\begin{figure}
\begin{centering}
\includegraphics[scale=0.55]{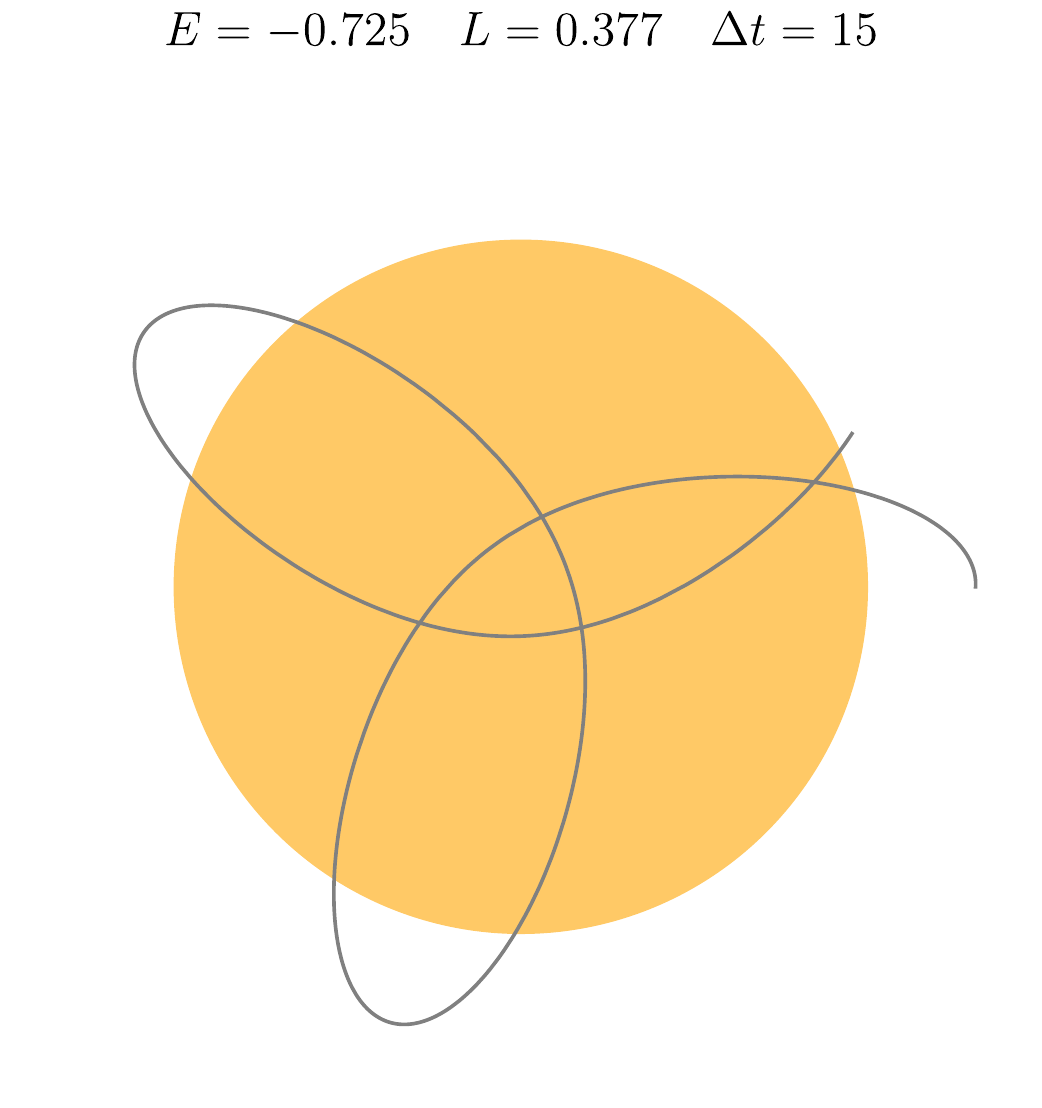}
\par\end{centering}

\protect\caption{\label{fig:TrajectoryDM}Trajectory of the DM particle for the orbit
$E=-0.725,\, L=0.377$ during a time interval $\Delta t=15$, starting at its apogee $r=1.31\,R_{\odot}$.  The yellow disk represents the Sun.}

\end{figure}

The Galilean invariant $\sigma\left(\left|\mathbf{w}-\mathbf{u}_{\mathcal{A}}\right|\right)$
can be obtained by integrating the differential cross section in Eq.~(\ref{eq:cross section}).
However, it should be noted that for DM mass around a few $\mathrm{GeV}$,
the typical momentum transfer in the thermal collision is of order
of $\mathrm{MeV}$, so we can neglect the Helm form factor for the
bound DM scattering process. Here we take operator $\hat{\mathcal{O}}_{11}$
as a specific example to illustrate how to calculate $\lambda$. First
it is not difficult to obtain the cross section

\begin{eqnarray}
\sigma_{11}\left(v_{\mathrm{rel}}\right) & = & \int_{0}^{2\mu_{A}v_{\mathrm{rel}}}\frac{c_{11}^{2}A^{2}}{2\pi v_{\mathrm{rel}}^{2}}P_{11}\left(v_{\mathrm{rel}}^{2},\, q^{2}\right)q\,\mathrm{d}q\nonumber \\
\nonumber \\
 & = & \mathrm{constant}\times v_{\mathrm{rel}}^{2},
\end{eqnarray}
with
\begin{equation}
\mathrm{constant}=\frac{4\, j_{\chi}\left(j_{\chi}+1\right)}{3}c_{11}^{2}\left(\frac{A^{2}\,\mu_{\mathcal{A}}^{4}}{2\pi\, m_{N}^{2}}\right),
\end{equation}
and then we input the $v_{\mathrm{rel}}^{2}$ reliance into integration
in Eq.~(\ref{eq:lambda}) as follows
\begin{eqnarray}
\lambda & = & \mathrm{constant}\times2\pi\int_{0}^{+\infty}\left[\int_{-1}^{1}\left(w^{2}+u_{\mathcal{A}}^{2}-2w\, u_{\mathcal{A}}\cdot x\right)^{3/2}\mathrm{d}x\right]\left(\sqrt{\pi}u_{0}\right)^{-3}\exp\left(-\frac{u_{\mathcal{A}}^{2}}{u_{0}^{2}}\right)u_{\mathcal{A}}^{2}\mathrm{d}u_{\mathcal{A}}\nonumber \\
\nonumber \\
 & = & \mathrm{constant}\times\frac{1}{4}\left[\frac{2\,\exp\left(-\frac{w^{2}}{u_{0}^{2}}\right)\, u_{0}\,\left(5\, u_{0}^{2}+2\, w^{2}\right)}{\sqrt{\pi}}+\left(\frac{3\, u_{0}^{4}}{w}+12\, u_{0}^{2}\, w+4\, w^{3}\right)\mathrm{erf}\left(\frac{w}{u_{0}}\right)\right].\nonumber \\
\nonumber \\
\end{eqnarray}
The analytic integration is performed using $\mathtt{Mathematica}$.

So once the DM particle motion is specified, the collision probability
can be evaluated explicitly with Eq.~(\ref{eq:collison P}). As an
illustration, a segment of the solar DM trajectory is shown in Fig.~\ref{fig:TrajectoryDM}. Similar depiction is presented in Ref.~\cite{Nussinov:2009ft}, where the bound orbit is calculated using an analytic approximation for the solar potential.

\section{\label{sec:appendixb}calculation of the scattering event rate}

\begin{figure}
\begin{centering}
\includegraphics[scale=0.35]{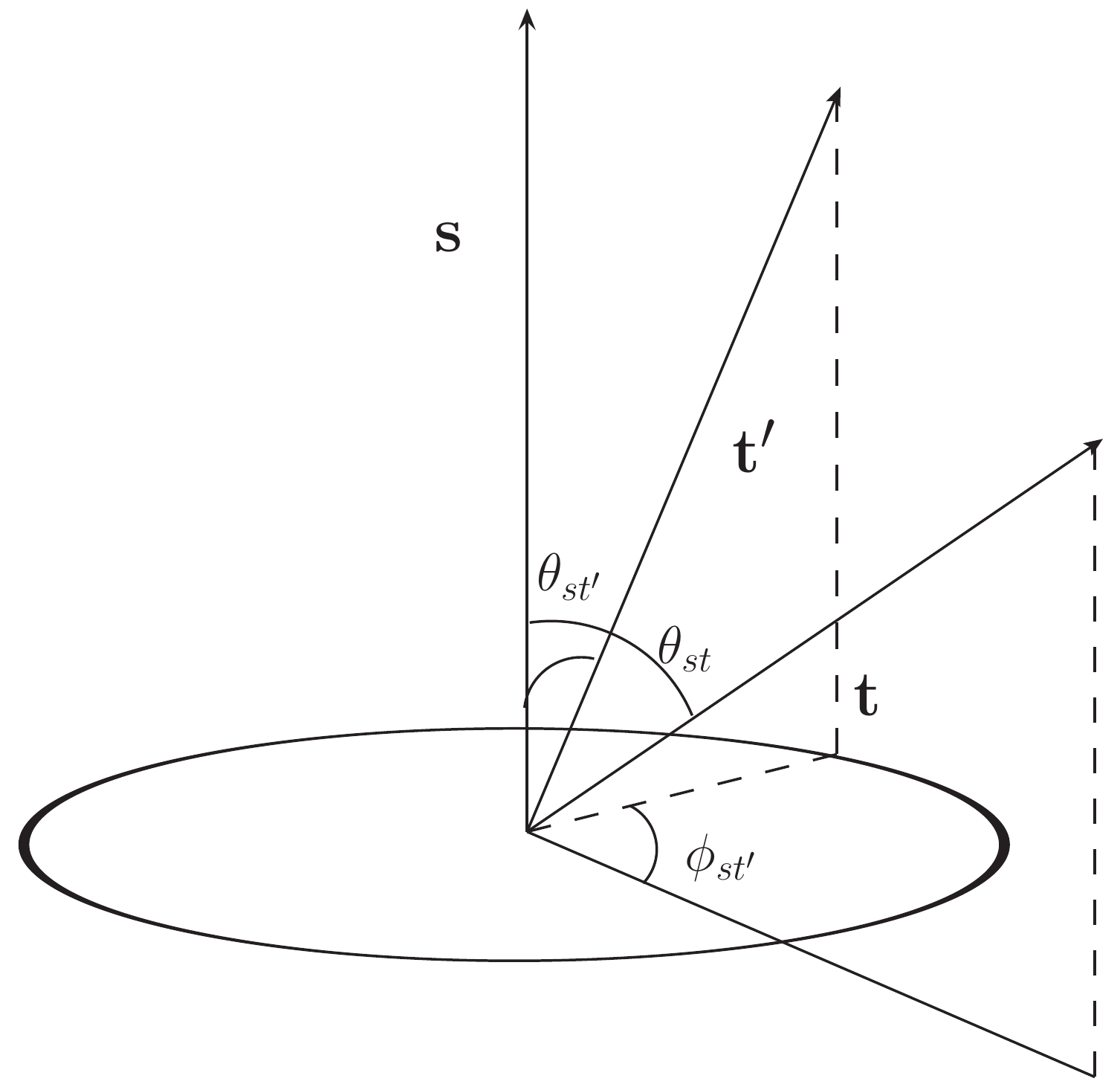}
\par\end{centering}

\protect\caption{\label{fig:fig2}The illustration of the vector $\mathbf{s},$ $\mathbf{t}$,
and $\mathbf{t}'$ with coordinates $\mathbf{s}=\left(0,\,0,\, s\right)$,
$\mathbf{t}=\left(t\sin\theta_{st},\,0,\, t\cos\theta_{st}\right)$
and $\mathbf{t}'=\left(t\sin\theta_{st'}\cos\phi_{st'},\, t\sin\theta_{st'}\sin\phi_{st'},\, t\cos\theta_{st'}\right)$,
respectively. Thus in Eq.~(\ref{eq:M1}) we have $\cos\theta_{t't}=\cos\theta_{st}\cos\theta_{st'}+\sin\theta_{st}\sin\theta_{st'}\cos\phi_{st'}$
and obtain Eq.~(\ref{eq:cosin}, \ref{eq:cosinprime}) by considering
$\mathbf{w}=\mathbf{s}+\mathbf{t}$, and $\mathbf{v}=\mathbf{s}+\mathbf{t}'$.}
\end{figure}

In this appendix we provide a detailed discussion on the scattering
event rate $R_{\mathcal{A}}\left(w\rightarrow v\right)$ at which
a DM particle scatters from initial velocity $w$ to final one $v$,
off a thermal bath composed of element $\mathcal{A}$ per unit volume.
Except for a few notations, our discussion follows closely the original
calculation in Refs.~\cite{Gould:1987ju,Gould:1989hm}. In short,
after a coordinate transformation from the solar system to the CM
system, Eq.~(\ref{eq:R1}) is expressed as an integration over the
transformed coordinates $\left(s,\, t\right)$ as the following:

\begin{eqnarray}
R_{\mathcal{A}}\left(w\rightarrow v\right) & =n_{\mathcal{A}} & (\eta_{\mathcal{A}}^{+})^{2}\, m_{\chi}^{2}\int f_{\mathcal{A}}\left(u_{\mathcal{A}}^{2}\right)\left\langle \mathcal{\left|M\right|}^{2}\right\rangle \left(w,\, v;\, s,\, t\right)\,\frac{v}{w}t\, ds\, dt\nonumber \\
 &  & \times\Theta\left(s+t-w\right)\Theta\left(w-\left|s-t\right|\right)\Theta\left(s+t-v\right)\Theta\left(v-\left|s-t\right|\right),\label{eq:R2}
\end{eqnarray}
where $\eta_{\mathcal{A}}^{+}\equiv1+\eta_{\mathcal{A}}\equiv1+m_{\chi}/m_{\mathcal{A}}$,
$\mathbf{s}=\left(m_{\chi}\mathbf{w}+m_{\mathcal{A}}\mathbf{u_{\mathcal{A}}}\right)/\left(m_{\mathcal{A}}+m_{\chi}\right)$
and $\mathbf{t}=m_{\mathcal{A}}\left(\mathbf{w}-\mathbf{u_{\mathcal{A}}}\right)/\left(m_{\mathcal{A}}+m_{\chi}\right)$
are the CM velocity and the DM incoming velocity in the CM frame,
respectively. $u_{\mathcal{A}}^{2}=\eta_{\mathcal{A}}^{+}s^{2}+\eta_{\mathcal{A}}\eta_{\mathcal{A}}^{+}t^{2}-\eta_{\mathcal{A}}w^{2}$,
$\mathcal{M}$ is the relevant scattering amplitude dependent on $\left(s,\, t\right)$
through the transferred momentum $\mathbf{q}=m_{\chi}\left(\mathbf{t}'-\mathbf{t}\right)$,
with $\mathbf{t}'$ the DM outgoing velocity in the CM frame, and $\Theta$
is the Heaviside step function. By illustrating
the relevant kinetic relation in Fig.~\ref{fig:fig2}, we express
the term $\left\langle \mathcal{\left|M\right|}^{2}\right\rangle $
as follows
\begin{eqnarray}
\left\langle \mathcal{\left|M\right|}^{2}\right\rangle \left(w,\, v;\, s,\, t\right) & = & \int_{0}^{2\pi}\left|\mathcal{M}\left(q^{2}\right)\right|^{2}\,\frac{d\phi_{st'}}{\left(2\pi\right)}\nonumber \\
 & = & \int_{0}^{2\pi}\left|\mathcal{M}\left(2\, m_{\chi}^{2}\, t^{2}\left[1-\cos\theta_{t't}\right]\right)\right|^{2}\,\frac{d\phi_{st'}}{\left(2\pi\right)}\nonumber \\
 & = & \int_{0}^{2\pi}\left|\mathcal{M}\left(2\, m_{\chi}^{2}\, t^{2}\left[1-\cos\theta_{st}\cos\theta_{st'}\right.\right.\right.\nonumber \\
 &  & \left.\left.\left.-\sin\theta_{st}\sin\theta_{st'}\cos\phi_{st'}\right]\,\right)\,\right|^{2}\,\frac{d\phi_{st'}}{\left(2\pi\right)},\label{eq:M1}
\end{eqnarray}
where
\begin{eqnarray}
\cos\theta_{st} & = & \frac{w^{2}-s^{2}-t^{2}}{2st},\label{eq:cosin}
\end{eqnarray}
and
\begin{eqnarray}
\cos\theta_{st'} & = & \frac{v^{2}-s^{2}-t^{2}}{2st}.\label{eq:cosinprime}
\end{eqnarray}
To integrate Eq.~(\ref{eq:R2}) we further change the variables as
the following
\begin{eqnarray}
x=t+s, &  & y=t-s,
\end{eqnarray}
or equivalently
\begin{eqnarray}
t=\frac{x+y}{2}, &  & s=\frac{x-y}{2},\label{eq:transfermation}
\end{eqnarray}

which also leads to the substitution for variables $\left(s,\, t\right)$
in the expression of $u_{\mathcal{A}}^{2}$:
\begin{eqnarray}
u_{\mathcal{A}}^{2} & = & \eta_{\mathcal{A}}^{+}\, s^{2}+\eta_{\mathcal{A}}\,\eta_{\mathcal{A}}^{+}\, t^{2}-\eta_{\mathcal{A}}\, w^{2}\nonumber \\
 & = & \left(\frac{\eta_{\mathcal{A}}^{+}\, x+\eta_{\mathcal{A}}^{-}\, y}{2}\right)^{2}+\eta_{\mathcal{A}}\,\left(y^{2}-w^{2}\right)\nonumber \\
 & = & \left(\frac{\eta_{\mathcal{A}}^{+}\, y+\eta_{\mathcal{A}}^{-}\, x}{2}\right)^{2}+\eta_{\mathcal{A}}\,\left(x^{2}-w^{2}\right),\label{eq:u square general}
\end{eqnarray}
with $\eta_{\mathcal{A}}^{-}=\eta_{\mathcal{A}}-1$, and
\begin{eqnarray}
\cos\theta_{st} & = & \frac{2w^{2}-x^{2}-y^{2}}{x^{2}-y^{2}},\label{eq:Cosine-st}
\end{eqnarray}

\begin{eqnarray}
\cos\theta_{st'} & = & \frac{2v^{2}-x^{2}-y^{2}}{x^{2}-y^{2}}.\label{eq:Cosine-stprime}
\end{eqnarray}
Applying these substitutions to Eq.~(\ref{eq:R2}) and assuming $v>w$
for evaporation, we have
\begin{eqnarray}
R_{\mathcal{A}}\left(w\rightarrow v\right) & = & n_{\mathcal{A}}\frac{(\eta_{\mathcal{A}}^{+})^{2}}{4}m_{\chi}^{2}\,\left(\sqrt{\pi}u_{0}\right)^{-3}\,\frac{v}{w}\,\int_{v}^{+\infty}dx\left[\int_{-w}^{+w}dy\,\left(x+y\right)\right.\nonumber \\
 &  & \left.\times\left\langle \mathcal{\left|M\right|}^{2}\right\rangle \left(w,\, v;\, x,\, y\right)\exp\left(-\frac{u_{\mathcal{A}}^{2}}{u_{0}^{2}}\right)\right],\label{eq:R3}
\end{eqnarray}
where
\begin{eqnarray}
\left\langle \mathcal{\left|M\right|}^{2}\right\rangle \left(w,\, v;\, x,\, y\right) & = & \int_{0}^{2\pi}\left|\mathcal{M}\left(2\, m_{\chi}^{2}\, t^{2}\left[1-\frac{\left(2w^{2}-x^{2}-y^{2}\right)\cdot\left(2v^{2}-x^{2}-y^{2}\right)}{\left[x^{2}-y^{2}\right]^{2}}\right.\right.\right.\nonumber \\
 &  & \left.\left.\left.-\frac{4\sqrt{\left(x^{2}-w^{2}\right)\cdot\left(x^{2}-v^{2}\right)}}{\left[x^{2}-y^{2}\right]^{2}}\left(w^{2}-y^{2}\right)\cos\phi_{st'}\right]\,\right)\,\right|^{2}\,\frac{d\phi_{st'}}{\left(2\pi\right)}.\nonumber \\
\end{eqnarray}
In practice, we simply numerically calculate Eq.~(\ref{eq:R3}) for
various DM-nucleon effective interactions, rather than finding an
analytic expression as has been done for the simplest case $\hat{\mathcal{O}}_{1}$
in Ref.~\cite{Gould:1987ju}.

\end{appendices}
\vspace{3cm}

\providecommand{\href}[2]{#2}\begingroup\raggedright\endgroup

\end{document}